# Hungry or not: how stellar-mass black holes grow (or don't) in dark matter mini-haloes at high resolution

Simone T. Gordon®,[1]★ Britton D. Smith,[1]★ Sadegh Khochfar[1] and John Anthony Regan®[2]

[1]*Institute for Astronomy, University of Edinburgh, Royal Observatory, Edinburgh EH9 3HJ, UK*
[2]*Centre for Astrophysics and Space Science Maynooth, Department of Theoretical Physics, Maynooth University, Maynooth, W23 F2H6, Ireland*

**ABSTRACT**
We compare the performance of the popular Bondi–Hoyle–Lyttleton accretion scheme with a simple mass-flux scheme applied to stellar-mass black holes (BHs) across six levels of increasing spatial resolution. Simulating the formation of BHs within cosmological mini-haloes at $z \sim 20$, we investigate scenarios both with and without supernova events, which result in BHs of initial mass 10.8 and 270 $M_\odot$, respectively. Our explicit focus on the stellar-mass range pushes the maximum resolution down to sub-$10^{-3}$ pc regimes, where more complicated gas dynamics are resolved. We observe efficient growth and rotationally supported, $\sim 10^{-1}$ pc-scale discs around all 270 $M_\odot$ BHs independent of resolution and accretion scheme, though clumps, bars, and spiral arm structures impact stability at high resolution. We analyse the effect of these instabilities on the accretion cycle. In contrast, all bar one of the 10.8 $M_\odot$ BHs fail to attract a disc and experience modest growth, even when characteristic scales of accretion and dynamical friction are reasonably resolved. While the two accretion schemes somewhat converge in mass growth for the 270 $M_\odot$ case over 1 Myr, the greater degree of gas fragmentation induces more randomness in the evolution of the 10.8 $M_\odot$ BHs. We conclude that early universe BHs of $M_{BH} \sim 10^1$ $M_\odot$ struggle to grow even in gas-rich environments without feedback in comparison to seeds of $M_{BH} \sim 10^2$ $M_\odot$, and the latter exhibit convergent growth histories across accretion schemes below a resolution of $dx = 1 \times 10^{-3}$ pc.

**Key words:** accretion, accretion discs – hydrodynamics – methods: numerical – software: simulations – dark ages, reionization, first stars – early Universe.

## 1 INTRODUCTION

The *JWST* and its predecessors have detected several supermassive black holes (SMBHs) exceeding $10^9$ $M_\odot$ as early as $z \sim 7$ (Mortlock et al. 2011; Mazzucchelli et al. 2017; Reed et al. 2019; Yang et al. 2020; Wang et al. 2021) and several possible Sagittarius A∗ sized SMBHs up to an unprecedented $z = 10.6$ (Goulding et al. 2023; Larson et al. 2023; Maiolino et al. 2024). These young behemoths pose a challenge to canonical theories of black hole (BH) evolution from the collapse of Population III (Pop III) stars (Pacucci et al. 2023) and have led to the emergence of three alternative formation channels: direct-collapse BHs from supermassive Pop III stars (Johnson et al. 2011; Agarwal et al. 2016; Regan & Downes 2018), stellar cluster mergers (Yajima & Khochfar 2016; Boekholt et al. 2018; Reinoso et al. 2023), and primordial BHs (Kawasaki, Kusenko & Yanagida 2012; Clesse & García-Bellido 2015; Carr & Silk 2018) (see Inayoshi, Visbal & Haiman 2020 for a recent review). Most of these seeding mechanisms are expected to begin during the cosmic dawn, at $z \sim 20$, or in the case of primordial BHs, even $z > 1000$. These epochs are likely beyond the reach of *JWST*, and therefore numerical simulations remain a vital tool in investigating the origin of SMBHs.

The extensive dynamic range traversed in cosmological and galaxy simulations demands significant computing power, which inevitably constrains spatial resolution. Sub-grid models are employed to approximate small-scale processes such as accretion and feedback using quantities measured at larger scales in cheaper, low-resolution simulations. The Bondi–Hoyle–Lyttleton (BHL) scheme has become a mainstream method for estimating accretion rates on to BHs, stars, and active galactic nuclei (Springel, Di Matteo & Hernquist 2005; Johansson, Naab & Burkert 2008; Smith et al. 2018; Comerford et al. 2019; Ehlert et al. 2023). It describes a compact body moving with relative velocity $v_\infty$ through a gas cloud of uniform density $\rho_{s,\infty}$ and sound speed $c_{s,\infty}$ measured 'at infinity' from the body, which accumulates mass from a stationary flow at the rate

$$\dot{M}_{BHL} = \alpha \frac{4\pi G^2 M_\bullet^2 \rho_\infty}{(c_{s,\infty}^2 + v_\infty^2)^{\frac{3}{2}}}, \quad (1)$$

where $\alpha$ is a dimensionless factor, $G$ is the gravitational constant, and $M_\bullet$ is the mass of the accretor ($M_\bullet = M_{BH}$ in this work) (Hoyle & Lyttleton 1939; Bondi 1952). The characteristic length-scales associated with this process are

$$R_{Bondi} = \frac{2GM_\bullet}{c_{s,\infty}^2}, \quad \text{when } c_{s,\infty}^2 > v_\infty^2 \quad (2)$$

★ E-mail: simone.gordon@ed.ac.uk (STG); britton.smith@ed.ac.uk (BDS)

**Table 1.** Literature summary of BHL model performance analysis. From left to right, the columns represent (1) the reference paper; (2) the simulation type, which can be an isolated model disc galaxy (*Galaxy 2D/3D*), a uniform mass resolution simulation of a cosmological volume (*Cosmo.*), a spherical distribution of gas (*Sphere*), or an idealised flow (*Bondi Flow*); (3) the numerical simulation scheme used, which can be one of moving-mesh (*MM*), smoothed particle-hydrodynamics (*SPH*), or adaptive mesh refinement (*AMR*); (4) the $\alpha$ parameter value adopted or the alternative model of accretion used; (5) inclusion of black hole feedback; (6) the minimum BH seed mass in units of solar masses; (7) the minimum halo mass in which the BH seeds form in units of solar mass; (8) the minimum cell width in the relevant simulations (not applicable in SPH case); and (9) the Bondi-radius resolution. This survey is not comprehensive and some values for the halo mass are only approximate.

| Study | Simulation | Type | $\alpha$ | BH feedback | $M_{BH}[M_\odot]$ | $M_{halo}[M_\odot]$ | $Min.\mathrm{d}x[pc]$ | $R_{Bondi}/\mathrm{d}x$ |
|---|---|---|---|---|---|---|---|---|
| Booth & Schaye (2009) | Cosmo. | SPH | $1^a$ | Yes | $1.2 \times 10^5$ | $5.7 \times 10^{10}$ | $>1^b$ | $<1$ |
| Curtis & Sijacki (2015) | Sphere | MM | 100 | No | $10^8$ | – | $10^{-3}$ | $>1$ |
| Negri & Volonteri (2017) | Galaxy 2D | Grid (no AMR) | 1 | Yes | $3 \times 10^7$ | $\simeq 3 \times 10^{10}$ | $10^{-1}$ | $\sim 1$ |
| Beckmann, Slyz & Devriendt (2018a) | Bondi Flow | AMR | 1 | No | – | – | $10^{-2}$ | 0.01–500 |
| Beckmann, Devriendt & Slyz (2018b) | Galaxy 3D | AMR | 1 | No | $10^1$–$10^5$ | $3 \times 10^{11}$ | $10^{-3}$ | $\sim 1$–$10^3$ |
| This study | Cosmo. | AMR | 1 | No | $10^1$–$10^2$ | $10^5$–$10^6$ | $10^{-4}$ | 0.2–$300^c$ |

*Notes.* $^a$The '$\beta$ –model' BHL variation was used with $\beta = 2$ (Booth & Schaye 2009).
$^b$The BH was $10^{-3}$ times the mass of the smallest gas particles and neither the Jeans mass nor the Bondi radius could be resolved, resulting in a low-resolution SPH simulation with equivalent grid cell resolution $\mathrm{d}x > 1$ pc.
$^c$This study measures resolution in $R_{HL} < R_{Bondi}$. The equivalent $R_{HL}$ resolution is $R_{HL}/\mathrm{d}x = 0.006$–6000.

and

$$R_{HL} = \frac{2GM_\bullet}{v_\infty^2}, \quad \text{when } c_{s,\infty}^2 < v_\infty^2, \tag{3}$$

referred to as the Bondi radius $R_{Bondi}$ and Hoyle–Lyttleton radius $R_{HL}$, respectively.

Despite its prevalence, the BHL model presents some issues when applied to more realistic environments. First, it is difficult to define gas properties 'at infinity' in a cosmological context. In practice, when $R_{Bondi/HL}$ is not resolved, $\rho_\infty$, $c_{s,\infty}$, and $v_\infty$ are often measured proximate to the BH particle (e.g. the host cell for grid codes or within the smoothing length for particle codes). However, complex local physics such as the self-gravity, turbulence, and thermodynamics of the gas along with accretion feedback are not accounted for in the BHL model nor captured accurately at low spatial resolution. The boost factor $\alpha \simeq 100$ was introduced by Springel, Di Matteo & Hernquist (2005) to compensate for the unresolved interstellar medium (ISM), thought to produce a net underestimate of the true accretion rate. Lacking the computational power to directly resolve the accretion region, the boost factor has been adopted by many authors who reason that increased gas resolution leads to higher densities and lower temperatures due to the activation of more cooling channels, resulting in cold, efficient accretion on to the BH (Robertson et al. 2006; Sijacki et al. 2007; Di Matteo et al. 2008; Johansson, Naab & Burkert 2008; Khalatyan et al. 2008; DeGraf et al. 2012; Hirschmann et al. 2014), though others opt for the non-boosted original if they deem the ISM to be sufficiently resolved (Smith et al. 2018; Dubois et al. 2021; Trebitsch et al. 2021). However, some have attempted to improve upon the BHL model by replacing $\alpha$ with a term parametrised by the temperature (Pelupessy, Di Matteo & Ciardi 2007), density (Booth & Schaye 2009), or vorticity (Curtis & Sijacki 2016) of the gas, or by simply adding an extra term that accounts for a specific mode of accretion (Grand et al. 2017). Others have built alternative models based on angular-momentum transport throughout the host galaxy (Debuhr, Quataert & Ma 2011; Hopkins & Quataert 2011; Rosas-Guevara et al. 2015). Certain contemporary simulation suites incorporate several of these accretion channels, each to be invoked under physical conditions that align with their underlying assumptions. The FIRE-2 simulations have a multitude of models; a simple fixed-rate per free-fall time, a torque-driven scheme in the vein of Hopkins & Quataert (2011), a constant-boost factor BHL-like model derived from isothermal spherical collapse (Shu 1977), and a shallow-sphere variation on that (Wellons et al. 2023). The ROMULUS suite employs a two-mode accretion scheme, amalgamating Booth & Schaye (2009)'s density-dependent boost factor with a BHL formula modified to use only the rotational component of the velocity under certain conditions to account for angular momentum (Tremmel et al. 2017). Nevertheless, other recent simulations continue to use the non-modified BHL scheme, such as OBELISK (Trebitsch et al. 2021) and NEWHORIZON (Dubois et al. 2021).

A number of studies have been carried out on the resolution dependence of the BHL model to assess the need for a boost factor, some of which are summarised in Table 1. Booth & Schaye (2009) emphasise the significance of Bondi radius resolution as distinct from the simulation resolution. $R_{Bondi}$ (equation 2) may become resolved for phases of the accretion cycle even in low-resolution simulations, during which the BHL formula with $\alpha > 1$ would overestimate the true accretion rate. They propose a modified boost factor: $\alpha = 1$ if $n < n_{th}$ and $\alpha = (n/n_{th})^\beta$ otherwise, where $n$ is the gas number density and $n_{th}$ is the star formation threshold density. They test a variety of $\beta$ values with the SPH code GADGET (Springel 2005) and find efficient accretion at $\beta = 2$ for a $\sim 10^5\,M_\odot$ SMBH at low resolution ($n_{th} = 0.1\,\mathrm{cm}^{-3}$) when neither the Jeans nor the Bondi mass was resolved. While they did not conduct a resolution dependence study and implicitly assumed gas becomes multiphase at $n_{th}$, they recognised the limits of a constant $\alpha$ and tried to incorporate the effects of resolution through their density-dependent boost factor. While their '$\beta$-model' found more success than other BHL-variations [it was adopted in Teyssier et al. (2011), Dubois, Volonteri & Silk (2014), and Volonteri et al. (2016)], they caution that the parameters may need to be tweaked for higher resolution simulations to achieve convergence with the observed scaling relations.

Curtis & Sijacki (2015) compare the mass-flux measured across a radius $r < R_{Bondi}$ from a $10^8\,M_\odot$ black hole held at the centre of a spherically symmetric distribution of gas in three simulations of increasing resolution. At the coarsest resolution, the locally measured mass-flux was lower than the 'theoretical' BHL value using gas attributes 'at infinity'. As resolution increased, the mass-flux and BHL values converged with reduced stochasticity over a $\sim 10^4$ yr period. Though the set-up is highly idealised, this experiment demonstrates how low-resolution simulations can give rise to artificially low densities around the BH and underestimate the accretion rate, an argument used by Booth & Schaye (2009) and others to justify the use of a fudge factor.

Negri & Volonteri (2017) extend the type of analysis performed by Curtis & Sijacki (2015) to a greater range of parameters and a more physically realistic environment. They compare Bondi accretion (no $v_\infty$ in the denominator of equation 1) with mass-flux accretion at 0.1 pc at three resolution levels on to an SMBH in a 2.5D galaxy simulation replete with mechanical and radiative active galactic nucleus (AGN) feedback. They make the type of weighting used on $\rho_\infty$ and $c_{s,\infty}$ and the radius $r_{\rm acc}$ at which these quantities are measured free parameters. In low-resolution simulations without feedback, they found that the BHL accretion rate is indeed underestimated with respect to the mass-flux rate. However, once feedback was introduced, the mass-weighted BHL-like scheme overestimated accretion at both low and high resolution and in low- and high-density environments. The difference was most pronounced in the high-resolution, high-density simulations which overestimated by an average of 10 times mass-flux rate over a 50 Myr period when the gas attributes were measured at $r_{\rm acc} = 3$ pc, a factor that grew to ∼100–1000 when $r_{\rm acc} = 30$ pc, 300 pc. This was largely due to the improved AGN feedback efficiency at high resolution that weakened as distance from the BH increased, allowing the simulations with large $r_{\rm acc}$ to grow very rapidly. The authors conclude that a boost factor (of undetermined value) could be justified in two scenarios: high-resolution, volume-weighted simulations that include AGN feedback, and low-resolution, mass-weighted simulations without AGN feedback. This work highlights the complicated interaction between feedback and accretion sub-grid models with each other and the resolution of the gas they depend on.

In a similar vein to both Negri & Volonteri (2017) and Curtis & Sijacki (2015), Beckmann, Slyz & Devriendt (2018a) (B18a henceforth) design a study of the BHL model's parameter space and its $R_{\rm Bondi}$-resolution dependence. Their set-up emulates the ideal conditions on which the BHL formula is based; a sink particle is embedded in a uniform gas flow of constant Mach number $\mathcal{M}$, which is toggled to investigate sub- and supersonic flow. They measure the simulation refinement level in units of $R_{\rm Bondi}$ grid-cell resolution $R_{\rm Bondi}/{\rm d}x$ and adjust this resolution by increasing or decreasing the BH mass ($R_{\rm Bondi} \propto M_{\rm BH}$). At low resolution ($R_{\rm Bondi}/{\rm d}x \leq 1$), the BHL rate calculated with gas attributes at 'at infinity' (the edge of the box) was close to the value using local gas properties; the BH does not exert a sufficient gravitational pull to affect the gas in its vicinity when its zone of influence is so underresolved. However, when $R_{\rm Bondi}/{\rm d}x > 100$ and $\mathcal{M} < 1.5$, the local BHL rate diverged from the 'at infinity' figure and converged to the lower local mass-flux rate. For adiabatic gas with $\mathcal{M} > 1.5$ at high resolution, gas instabilities appeared and started to dominate the flow, resulting in diminished local accretion rates that reached just 10 per cent of the 'at infinity' BHL figure. In their highly idealised set-up, the local BHL and mass-flux accretion schemes are robust against complex flow configurations and increasing resolution as they produce convergent solutions.

In Beckmann, Devriendt & Slyz (2018b) (B18b henceforth), the investigation into the impact of resolution on the BHL model is applied to the more realistic set-up of an isolated galaxy at the centre of which is placed a single BH from a range of masses. They first demonstrate that the local BHL rate converges to the mass-flux rate at sufficiently high resolution, reinforcing the results from B18a. When rotation is added to the galaxy and the resolution of the accretion region is gradually increased to ${\rm d}x = 0.015$ pc, a parsec-scale disc quickly develops around BHs with masses $M_{\rm BH} > 10^5 \, {\rm M}_\odot$, feeding their rapid growth. Such discs also appear near smaller BHs, though the particle does not remain embedded in the centre of the gas cloud as it collapses and does not grow efficiently as a result. The choice of seed mass (and therefore $R_{\rm Bondi/HL}$ resolution) led to starkly different growth histories, with the largest growing by up to two orders of magnitude and the smallest showing no appreciable growth. Once the small BHs were fixed to the centre of the gas via a 'maximum drag-force algorithm', they indeed quickly converged with the growth of the larger BHs. Another set of simulations was run at varied spatial resolutions with the initial BH mass chosen so that $R_{\rm Bondi,\, init} \simeq {\rm d}x$. While all jump to ∼ $10^5 \, {\rm M}_\odot$ in < 0.1 Myr, despite starting from orders of magnitude different seed masses, the subsequent gas properties and accretion patterns vary such that mass convergence is not achieved by $t = 5$ Myr. The accretion cycle in high-resolution simulations is episodic, driven by infalling clumps of gas that form in the spiral arms of its nuclear disc. Low-resolution simulations experience chaotic accretion that fluctuates by orders of magnitude on short time-scales. These simulations reveal that even in the absence of feedback, the impact of resolution on gas properties and the BHL accretion cycle is significant, while the memory of the seed mass is quickly washed out by an initial burst of growth fed by a gas-rich environment, so long as the particle can remain bound to the cloud as it collapses.

This work presents the first resolution study of BHL accretion on to stellar-mass black hole remnants of Pop III stars situated in high-redshift dark matter mini-haloes formed with cosmological initial conditions and metal-free gas. Most prior works on BH accretion in galactic environments use $> 10^5 \, {\rm M}_\odot$ seed masses (e.g. Negri & Volonteri 2017, and other works discussed above), though SMBH super-Eddington accretion has potentially been observed as early as $z = 10.6$ (Maiolino et al. 2024), keeping open the possibility for $z \sim 25$ (or even primordial) stellar-mass seeds. As such, accurately simulating the evolution of these BHs and understanding the impact of spatial resolution remains a relevant challenge. Wielding the power of adaptive mesh refinement (AMR), we developed a BH refinement algorithm that ensures the accretion region remains at a fixed resolution level set by $R_{\rm HL}$, which we resolve by up to 16 grid cells. Some simulations reach a spatial resolution of ${\rm d}x = 8 \times 10^{-4}$ pc, offering an in-depth examination of the complex gas dynamics within the BHs' gravitational zone of influence. Table 2 provides a glossary of commonly used variables in this work, for the reader's reference.

This paper is organised as follows. Section 2 details the set-up of our simulations and the modifications and additions we made to the code. The results in Section 3 focus on a 270 $\rm M_\odot$ BH formed from the collapse of a Pop III star of equal mass without a supernova and Section 3.4 covers the evolution of a 10.8 $\rm M_\odot$ BH formed from a 40 $\rm M_\odot$ Pop III star both with and without a Type II supernova. Section 4 contains a comprehensive discussion of the results and comparisons with other works. Finally, we summarise our conclusions in Section 5.

## 2 THE SIMULATIONS

We use the simulation code ENZO (Bryan & Enzo Collaboration 2014) to generate all data analysed in this work. ENZO is a hydrodynamical, block-structured AMR + $N$-body cosmological simulation code that is capable of achieving very high spatial and temporal resolution. Our simulations are designed to follow the collapse of a metal-free gas cloud within a single cosmological halo. We initialise the simulations with a 500 kpc h$^{-1}$ comoving box at $z = 180$ using the MUSIC (Hahn & Abel 2011) initial conditions generator with the *WMAP 7* best-fitting cosmological parameters, $\Omega_{\rm m} = 0.266$, $\Omega_\lambda = 0.732$, $\Omega_{\rm b} = 0.0449$, $H_0 = 71.0 \, {\rm km \, s^{-1} Mpc^{-1}}$, $\sigma_8 = 0.801$, and $n_s = 0.963$ (Komatsu et al. 2011), the Hu & Eisenstein (1999) transfer function, and second-order Lagrangian perturbation theory. We run

**Table 2.** Definitions of commonly used variables in this work.

| Variable | Definition |
|---|---|
| $dx$ | Cell width of the high-resolution grid |
| $\dot{M}_{BHL}$ | BHL accretion rate (equation 1) |
| $R_{Bondi}$ | Bondi radius (equation 2) |
| $R_{HL}$ | Hoyle–Lyttleton radius (equation 3) |
| $\rho_\infty$ | Gas density in accretion region |
| $c_{s,\infty}$ | Sound speed in accretion region |
| $v_\infty$ | Relative velocity of the gas in accretion region with respect to the BH |
| $\dot{M}_{flux}$ | Mass-flux accretion rate (equation 11) |
| $r_K$ | Kernel radius (equation 12) |
| $n$ | Total hydrogen nuclei number density |
| $r_{BHL}$ | BHL interpolation radius (equation 13) |
| $l_{cool}$ | Cooling length |
| $R_{bar}$ | Radius of a bar-like instability |
| $R_{co\text{-}rot}$ | Co-rotational radius |
| $Q$ | Toomre Q instability parameter |
| $Q_{adv}$ | Advective cooling rate (equation 14) |
| $Q_{rad}$ | Radiative cooling rate |

**Table 3.** Summary of the environment in which the BH forms for each set of initial conditions. Any 'average' value is measured over the BH accretion region (see Section 2.3 for the definition).

| Conditions at time of BH formation | Halo 1 | Halo 2 |
|---|---|---|
| Formation time [Myr] | 124.76 | 195.59 |
| Redshift | 26.31 | 19.23 |
| Refinement level | 14 | 15 |
| Minimum $dx$ [com. pc] | 0.327 | 0.168 |
| Virial mass $M_{200}$ [$M_\odot$] | $2.82 \times 10^5$ | $1.29 \times 10^6$ |
| Virial radius $R_{200}$ [pc] | 82.34 | 187.10 |
| Average $n$ [cm$^{-3}$] | $5.59 \times 10^{05}$ | $5.15 \times 10^5$ |
| Average temperature [K] | 496.97 | 409.92 |
| Average $c_{s,\infty}$ [km s$^{-1}$] | 0.51 | 0.34 |
| Average $v_\infty$ [km s$^{-1}$] | 1.96 | 1.98 |

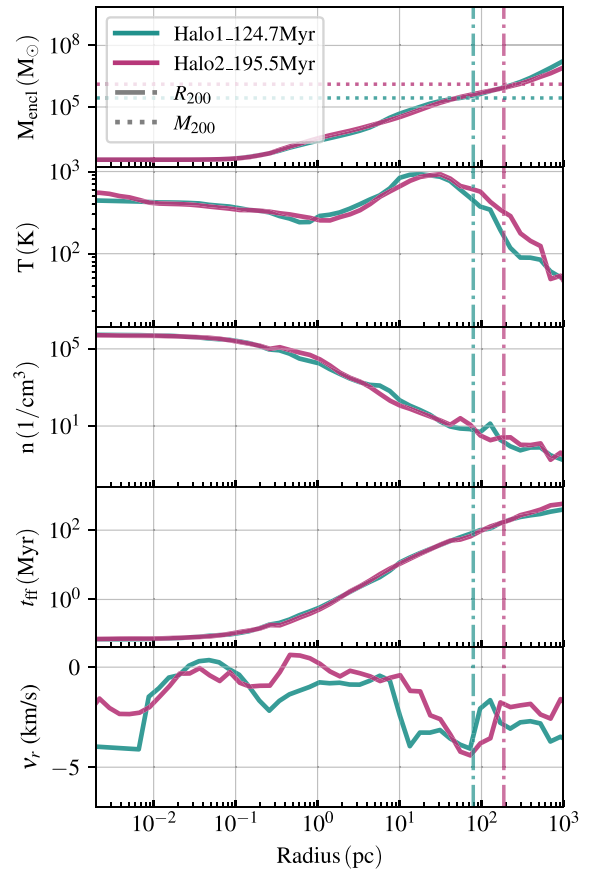

**Figure 1.** Radial profiles of halo properties centred on the BH particle at the time of formation. From top to bottom, the panels display the total enclosed gas mass, mass-weighted temperature, mass-weighted total number density, free-fall time, and radial velocity. The virial radius $R_{200}$ and virial mass $M_{200}$ are denoted by *dash–dot* and *dashed* lines, respectively. Both haloes lie under the atomic cooling mass limit and contain a $6 \times 10^2 \, M_\odot$ core with a radius of 0.2 pc.

all simulations with a root grid of $128^3$ dark matter particles and cells.

In order to identify a suitable halo for BH formation, we run dark-matter-only simulations to $z = 10$ and locate a halo with total mass $\simeq 10^7 \, M_\odot$ in two sets of initial conditions using the ROCKSTAR halo finder (Behroozi et al. 2013). We then re-simulate from $z = 180$ with an additional four levels of telescoping refinement around the target halo, adding baryons in the last iteration. The high-resolution region is a rectangular prism that contains all dark matter particles destined to reside within three virial radii of the target halo at $z = 10$. These particles are designated as 'must-refine particles,' ensuring AMR is exclusively applied within this zone (Simpson et al. 2013). Prior to AMR, the high-resolution region has an effective resolution of $2048^3$, corresponding to a comoving spatial resolution of 0.244 kpc h$^{-1}$, a baryon mass resolution of 0.259 $M_\odot$, and a dark matter mass resolution of 1.274 $M_\odot$.

A summary of the pre-BH accretion properties of each halo is shown in Table 3 and Fig. 1. Halo 1 forms at the higher end of the expected redshift for the onset of Pop III star formation, while Halo 2 forms at the lower end. Both are close to the minimum baryonic mass a virialised cloud must have to cool sufficiently for the formation of Pop III stars (Tegmark et al. 1997; Abel, Bryan & Norman 2002). This distinguishes these structures as mini-haloes as opposed to young galaxies (Greif et al. 2008).

The analysis in Section 3 has been limited to the first 1 Myr of BH evolution. Some of our simulations have very high spatial resolution, which is computationally expensive. However, 1 Myr corresponds to the $t_{ff}$ within the central $r = 1$ pc of the mini-halo and it can be seen from Fig. 1 that at this point the mass-weighted density starts to decline rapidly and the enclosed gas mass levels off. Hence, this cut-off is an appropriate choice to study the short-term mass accretion from this local environment.

### 2.1 Cooling and heating

We use the GRACKLE non-equilibrium chemistry network (Smith et al. 2017) which follows the interactions of the nine dominant chemical species in primordial gas: H, H$^+$, H$^-$, $e^-$, He, He$^+$, He$^{++}$, H$_2$, and H$_2^+$. The radiative losses from atomic and molecular line cooling, Compton cooling, and heating of free electrons by cosmic microwave background photons are appropriately treated in the optically thin limit. This network includes H$_2$ formation from the H$^-$ and H$_2^+$ channels, three-body formation according to the Glover (2008) prescription, H$_2$ rotational transitions, chemical heating, and collision-induced emission [important in $n \geq 10^{14}$ cm$^{-3}$ gas (Ripamonti & Abel 2004)]. The simulations presented here reach densities of $n > 10^9$ cm$^{-3}$ in which regime the three-body H$_2$ formation channel has been shown to dominate the cooling (Abel, Bryan & Norman 2002; Turk et al. 2010); hence, its inclusion is essential to accurately capture the gas evolution on these scales. Metals are introduced in a subset of our simulations via Type II supernova events. We use the spectral synthesis software CLOUDY (Ferland et al. 2013) to apply metal cooling rates under collisional ionization equilibrium (no incident radiation field). These cooling rates are valid for gas densities of $10^{-6}\,\mathrm{cm}^{-3} < n < 10^{12}\,\mathrm{cm}^{-3}$, metallicities of $10^{-6}\,\mathrm{Z}_\odot < Z < 10\,\mathrm{Z}_\odot$, and temperatures of $10\,\mathrm{K} < T < 10^8\,\mathrm{K}$.

Cooling from HD is not included in this model. While HD dominates over $H_2$ at $T < 100$ K, collisional ionization in primordial gas is required to form the molecule in high abundance. This can occur in shock heating to temperatures above $10^4$ K, as in the virialization of atomic cooling haloes, or through the radiation from nearby Pop III stars (Greif et al. 2010). The BHs in this work are assumed to form from 'the first star in the Universe' [a so-called 'Pop III.I' star (Tan & McKee 2008)] in low-mass mini-haloes in which the HD fraction is not high enough to cool the gas to $\sim T_{\mathrm{CMB}}$ (Yoshida et al. 2007). Therefore, this omission from the chemistry model is unlikely to substantially affect the outcome of these simulations.

## 2.2 SmartStars

We use the ENZO module SmartStars to simulate particle formation and accretion within the high-resolution dark matter halo. This module was first created by Regan & Downes (2018) to simulate the evolution of supermassive Pop III stars in atomic cooling haloes and their subsequent collapse into black holes. It forms part of the ActiveParticles class which was built to have robust feedback capabilities, including radiative, mechanical, and thermal modes for stars and BHs alike. The flexibility of the module, in particular the ability of a SmartStar to change particle type (e.g. from a Pop III star particle to a BH sink particle), made it an attractive option for our work as it leaves the door open on additional physics to extend our study to more realistic environments. However, it was necessary to adapt the code for four processes to make it suitable for our current investigation:

(i) AMR
(ii) Pop III star formation
(iii) BH formation
(iv) BH accretion

The following subsections will detail these code adaptations and extensions.

## 2.3 Adaptive mesh refinement

AMR occurs only in the high-resolution zoom-in region. Grid cells are split by factor of 2 in each dimension when certain conditions are met. These simulations use four criteria native to ENZO:

(i) the cell contains a 'must-refine particle', indicating that it is destined to reside in the target halo at the end of the simulation.
(ii) the dark matter mass within a grid cell is greater than four times the initial mass (i.e. when more than four dark matter particles are in the same cell).
(iii) the gas mass within a grid cell is greater than four times the mean baryon mass per cell on the root grid multiplied by a factor, $2^{-0.2L}$, where L is the refinement level.
(iv) the local Jeans length is resolved by less than 32 or 64 cells.

While it has been shown that resolving the local Jeans length by at least 64 cells [a factor of 16 higher than the criterion put forth by Truelove et al. (1997)] is important to guarantee that fragmentation is purely physical in origin (Federrath et al. 2011; Meece, Smith & O'Shea 2014), it became necessary to relax this to 32 cells to achieve the desired level of refinement in some lower resolution simulations. In addition to these pre-existing conditions, we added another to control the resolution of the physics pertinent to accretion:

(v) the local scale radius $r_{\mathrm{scale}} = \max(R_{\mathrm{HL}}, R_{\mathrm{Bondi}})$ is not resolved within a user-specified number of cell widths.

The scale radius $r_{\mathrm{scale}}$ is taken to be the maximum of equation (2) and equation (3) where $c_{s,\infty}^2$ and $v_\infty^2$ are calculated by sampling cell properties within a kernel radius $r_K$ around the BH and taking a mass- and kernel-weighted average of these values. The radius weighting is applied on each cell $i$ through the Gaussian kernel

$$\omega_i = \exp(-r_i^2/r_K^2), \qquad (4)$$

where $r_i < r_K$ is the distance from the centre of a neighbouring cell to the BH (as per B18b). The kernel radius $r_K$ is defined differently in the underresolved and overresolved cases and will be covered in Section 2.5. Hence, the properties $c_{s,\infty}$ and $v_\infty$ (denoted $x_i$) are calculated by

$$x_{\mathrm{BHL}} = \frac{\sum_i x_i m_i \omega_i}{\sum_i m_i \omega_i}, \qquad (5)$$

where $m_i$ is the mass of the cell and $\omega_i$ is the Gaussian kernel of equation (4). Refinement is triggered when

$$\mathrm{d}x > r_{\mathrm{scale}}/\mathrm{BF}, \qquad (6)$$

where BF = BHLRadiusRefinementFactor[1] is a user-set parameter defining the number of cells by which the scale radius should be resolved. When BF > 1, the scale radius is resolved within one or more cells. In contrast, a value of 0 < BF < 1 indicates that the scale radius is underresolved, covering only a fraction of a cell's width. This condition is evaluated at every time-step. The implementation is biased towards overresolving rather than underresolving. For instance, if the scale radius is currently resolved by 0.9 cells and BF = 1, refinement is triggered and the radius is resolved by 1.8 cells in subsequent steps.

The minimum size of the maximally refined grid depends on the resolution:

$$N_{\mathrm{grid}} = (2\max((r_{\mathrm{scale}}/\mathrm{d}x) + 2, 5))^3. \qquad (7)$$

For example, if $r_{\mathrm{scale}}$ is resolved by eight grid cells, then $N_{\mathrm{grid}} = (2 \times 10)^3 = 20^3$ cells. Alternatively, if $r_{\mathrm{scale}}$ is underresolved and only covers one-half of a grid cell, $N_{\mathrm{grid}}$ will take the minimum value of $(2 \times 5)^3$ cells. This is to ensure that we are always sampling gas properties and removing gas from cells at the maximum refinement level on which the BH lives. It was found in Section 3.2 of B18b that this is necessary for convergence across the gas in density and sound speed as well as the accretion rate of a $260\,\mathrm{M}_\odot$ BH. In practice, the high-resolution grids in our simulations generally cover a region much larger than equation (7) due to the combination of refinement criteria. Fig. 2 shows a density projection of one of our simulations centred on the BH particle with the maximum resolution grids overlaid. The accretion region (marked by a white circle) is clearly covered, as are the dense spiral arms and clumps extending past $r = 0.15$ pc.

Furthermore, we modified the SmartStars module in a more fundamental manner by removing the requirement that the particle lives on the user-set maximum refinement level. This was necessary to enable the scale-radius criterion to lead to the refinement and de-refinement of the accretion region. It also has additional computational expense benefits in certain environments.

---

[1]This is the exact name of the parameter used in the code.

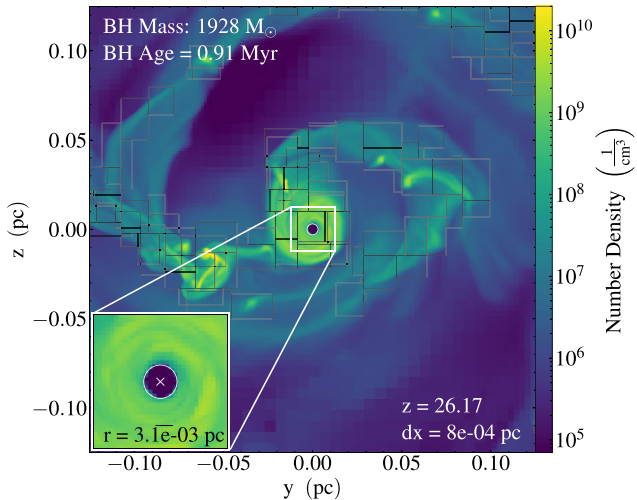

**Figure 2.** Total hydrogen nuclei number density projection of simulation `1B.m16` (see Table 4 for details) overlaid with grid boundaries (*black boxes*) at the maximum refinement level of 18, corresponding to a minimum cell width of $dx = 8 \times 10^{-4}$ pc. The position of the BH is marked with a *white cross* in the zoomed-in inset box around the accretion region, denoted by a *white circle*. The high refinement grids amply cover the accretion region as well as the dense spiral arms.

### 2.4 Star/black hole particle formation

We simulate the formation of Schwarzschild (non-spinning) BHs from Pop III stars. However, we do not simulate the main-sequence phase of the progenitor stars to approximate the recent finding that ionizing radiation remains trapped in the dense stellar accretion discs at high resolution (Jaura et al. 2022). A Pop III star particle is formed in a grid cell where the following conditions are met:

(i) the proper baryon number density exceeds $10^6$ cm$^{-3}$.
(ii) the gas flow is convergent.
(iii) the molecular hydrogen mass fraction $f_{H2} \equiv (\rho_{H2} + \rho_{H2+})/\rho_b$ exceeds $5 \times 10^{-4}$, where $\rho_{H2}$, $\rho_{H2+}$, and $\rho_b$ are the neutral molecular hydrogen, singly ionised molecular hydrogen, and the total baryon densities, respectively.
(iv) there is at least twice the mass of the particle in gas within a sphere of radius $r \simeq 1$ pc.

The standard `SmartStars` stellar seeding mechanism uses radiative lifetime as the user-set fixed parameter; the star is inserted with an arbitrarily small mass and then accretes to attain an environment-dependent final mass (Regan & Downes 2018). However, to control the initial mass of the BH, we required a pre-determined final mass for the star. We implemented a formation model based on Wise et al. (2012) in which the star is initialised with its final mass as specified by the user, and stellar accretion is turned off. This approach necessitates the fourth formation criterion, which ensures the conservation of energy–momentum by instantaneously removing the star's entire mass from the grid. A 'searching sphere' is incrementally extended around the particle until it encapsulates twice the mass of the star, at which point the particle mass is removed from all cells within the sphere. The mass limit of twice the star particle mass was chosen so no more than 50 per cent of the gas is removed from any cell to avoid numerical issues arising from excessive instantaneous gas removal. While the main-sequence lifetime associated with the star's mass typically follows values from table 4 in Schaerer (2002), we set it to zero in our study, allowing for immediate BH formation.

A zero metallicity environment results in less efficient stellar winds (Gräfener & Hamann 2008; Rickard et al. 2022), which in turn leads to a less pronounced mass-loss during the final stages of the life of the star. Thus, Pop III stars might build massive helium cores, which is not possible for their high-metallicity counterparts. The mass of the helium core is a key factor in determining the initial mass of the BH that forms in its wake. For the 40 M$_\odot$ stars, we assume the Heger–Woosley relation (Woosley, Heger & Weaver 2002) to calculate the helium core, which is 10.8 M$_\odot$. The mass-loss is converted to kinetic energy in the Type II core-collapse supernova. For the 270 M$_\odot$ direct-collapse case, we assume negligible mass-loss, hence the helium core remains at 270 M$_\odot$, and there is no supernova event before a BH of equal mass is formed.

### 2.5 Black hole accretion

Once the BH has been refined to the level dictated by the four criteria in Section 2.3, it is allowed to accrete gas from within the accretion radius (a minimum of 1 cell width). We employ two accretion schemes in our study, $\dot M_{\rm BHL}$ and $\dot M_{\rm flux}$. The differences in the schemes are detailed in the following two subsections, but there are some common features to both. First, there is a 75 per cent cap on the mass of gas that can be removed from a cell of mass $m_i$ at each time-step $dt$. Hence, the total mass removed per time-step $\Delta M_{\rm BH}$ is

$$\Delta M_{\rm BH} = \sum_i \min\left(\frac{dt\,\dot M_{\rm BHL/flux}\,m_i\omega_i}{(\Sigma_j m_j\omega_j)},\,0.75\,m_i\right), \quad (8)$$

where $\omega_i$ is the Gaussian kernel of equation (4) and the denominator is the sum of all cells in the accretion sphere, as in equation (5). The BH mass is updated accordingly and the accretion rate is stored every $\simeq 100$ yr as

$$\dot M_{\rm BH} = \left.\frac{\Delta M_{\rm BH}}{dt}\right|_{dt \simeq 100\,\rm yr}. \quad (9)$$

Furthermore, we do not limit accretion by the Eddington rate

$$\dot M_{\rm Edd} = \frac{4\pi G M_{\rm BH} m_{\rm p}}{\epsilon_{\rm r} \sigma_{\rm T} c}, \quad (10)$$

where $\epsilon_{\rm r}$ is the radiative efficiency, $\sigma_{\rm T}$ the Thompson cross-section, $m_{\rm p}$ the proton mass, and $c$ the speed of light in a vacuum.

#### 2.5.1 Mass-flux accretion

The mass-flux scheme is a simple accretion model that only depends on the quantity of in-falling matter across a surface at a radius of $r_{\rm surf}$ from the BH:

$$\dot M_{\rm flux} = 4\pi \int_{r_{\rm surf}-dx}^{r_{\rm surf}} dr\,\rho\,r^2 v^-(r), \quad (11)$$

where $\rho$ and $v^-(r)$ are the density and radial velocity of a cell intersecting the surface, respectively, where only cells with negative radial velocity (in-falling) are included. The surface width, $dx$, is the high-resolution cell width, and $r_{\rm surf} = 4dx$. The latter is chosen as a compromise between sampling the cells closest to the BH at each resolution scale and retaining some spherical symmetry (as in Regan et al. 2019). We include all cells which fall in the range $r_{\rm surf} - dx < r < r_{\rm surf}$ in the calculation of equation (11). It is worth emphasizing that not all mass in the cells which fulfil the prior criteria is absorbed by the BH. Due to the 75 per cent cap on the mass that can be removed per time-step, the actual accreted mass $\Delta M_{BH}$ is limited by the local supply of gas, and, in practice, is often less than $\dot{M}_{flux} \, dt$, as per equation (8). Finally, we set $r_K = r_{surf}$, where $r_K$ is the kernel radius used to calculate $\omega_i$ in equation (8).

**Table 4.** Summary of simulations of the 270 M$_\odot$ BH across two sets of initial conditions and accretion schemes at various levels of refinement. Each simulation is labelled according to the following convention: the host halo (1 or 2), the initial size of the BH ('B' for the 'big' 270 M$_\odot$ or 'S' for the 'small' 10.8 M$_\odot$ BH), the accretion scheme used ('b' for 'BHL' or 'm' for 'Mass-Flux'), and the spatial resolution measured in units of the initial Hoyle–Lyttleton radius resolution, $R_{HL}/dx$. For example, `1B.m04` refers to a simulation of a Halo 1, 270 M$_\odot$ BH with $R_{HL}$ resolved by four cell widths, whereas `1B.mf4` refers to a lower resolution simulation with $R_{HL}$ underresolved by one-fourth of a cell width (where the 'f' stands for 'fraction'). From left to right, the columns represent (1) the simulation name; (2) the mass of the BH in solar masses; (3) the accretion scheme used, which can be either the BHL scheme (equation 1) or mass-flux (equation 11); (4) the number of cells by which the initial scale radius is resolved; (5) the resolution level of the ENZO simulation; (6) the minimum cell width in units of proper parsecs; and (7) the minimum size of the high-resolution accretion region in number of cells.

| | | | 270 M$_\odot$ BH simulations | | | |
|---|---|---|---|---|---|---|
| Name | Mass [M$_\odot$] | Accretion | $R_{HL}$ resolution | Maximum refinement level | dx [pc] | $N_{grid}$ [cells] |
| 1B.b01 | 270 | BHL | 1 | 14 | $1.23 \times 10^{-02}$ | $5^3$ |
| 1B.b04 | 270 | BHL | 4 | 16 | $3.07 \times 10^{-03}$ | $6^3$ |
| 1B.b08 | 270 | BHL | 8 | 17 | $1.54 \times 10^{-03}$ | $11^3$ |
| 1B.b16 | 270 | BHL | 16 | 18 | $7.69 \times 10^{-04}$ | $20^3$ |
| 1B.m01 | 270 | Mass-Flux | 1 | 14 | $1.23 \times 10^{-02}$ | $5^3$ |
| 1B.m04 | 270 | Mass-Flux | 4 | 16 | $3.07 \times 10^{-03}$ | $6^3$ |
| 1B.m08 | 270 | Mass-Flux | 8 | 17 | $1.54 \times 10^{-03}$ | $11^3$ |
| 1B.m16 | 270 | Mass-Flux | 16 | 18 | $7.69 \times 10^{-04}$ | $20^3$ |
| 2B.b01 | 270 | BHL | 1 | 15 | $8.30 \times 10^{-03}$ | $5^3$ |
| 2B.b04 | 270 | BHL | 4 | 17 | $2.07 \times 10^{-03}$ | $6^3$ |
| 2B.b08 | 270 | BHL | 8 | 18 | $1.04 \times 10^{-03}$ | $11^3$ |
| 2B.b16 | 270 | BHL | 16 | 19 | $5.18 \times 10^{-04}$ | $19^3$ |
| 2B.m01 | 270 | Mass-Flux | 1 | 15 | $8.30 \times 10^{-03}$ | $5^3$ |
| 2B.m04 | 270 | Mass-Flux | 4 | 17 | $2.07 \times 10^{-03}$ | $6^3$ |
| 2B.m08 | 270 | Mass-Flux | 8 | 18 | $1.04 \times 10^{-03}$ | $11^3$ |
| 2B.m16 | 270 | Mass-Flux | 16 | 19 | $5.18 \times 10^{-04}$ | $19^3$ |

### 2.5.2 BHL accretion

We also apply the canonical BHL model of equation (1). The equation input variables $\rho_\infty$, $c_{s,\infty}$, and $v_\infty$ are calculated with equation (5). We calculate the kernel radius by considering the resolution of the interpolated BHL radius

$$r_K = \begin{cases} dx_{min}, & \text{if } r_{BHL} \leq dx_{min} \\ r_{BHL}, & \text{if } dx_{min} < r_{BHL} \leq 2\,dx_{min} \\ 2\,dx_{min} & \text{if } r_{BHL} > 2\,dx_{min} \end{cases} \quad (12)$$

depending on the size of the interpolated BHL radius

$$r_{BHL} = \frac{G M_{BH}}{c_{s,\infty}^2 + v_\infty^2}, \quad (13)$$

which is calculated on the fly using the BH mass $M_{BH}$, the sound speed, and the relative velocity of the gas in the cell hosting the BH particle, $c_{s,\infty}$ and $v_\infty$ (see the Section 1 for an explanation of this counter-intuitive notation).

## 3 RESULTS

We first analyse the impact of simulation resolution, accretion scheme, and environment on the evolution of a 270 M$_\odot$ BH. According to the Heger–Woosley relation between BHs and their progenitor Pop III stars (Woosley, Heger & Weaver 2002), a 270 M$_\odot$ metal-free star collapses directly into a BH of approximately equal mass. Since we do not simulate the main-sequence lifetime of the star and there is no supernova event preceding its formation, this BH is born into an environment of relatively dense pristine gas. We first characterise the accretion cycle and growth of the more massive BH (Section 3.1), followed by the properties of the gas in its vicinity (Section 3.2), and instabilities that appear in its host disc (Section 3.3). All simulations pertaining to these sections are shown in Table 4. We then compare this analysis to simulations of a 10.8 M$_\odot$ BH remnant of a 40 M$_\odot$ Pop III star (Section 3.4), which forms following a $10^{51}$ erg-Type II core-collapse supernova (also according to the Heger–Woosley relation).

### 3.1 BH accretion cycle and growth

#### 3.1.1 At baseline resolution

In both sets of haloes and accretion schemes at baseline resolution (`m01` and `b01` – see the caption of Table 4 for an explanation of the simulation labelling convention), the 270 M$_\odot$ BH grows to 2000–3000 M$_\odot$ over a period of 1 Myr. Despite the similarity in the final masses, there are considerable variations in the accretion history between the two haloes. Referring to Fig. 3, Halo 1 BHs display rapid growth in the initial 0.2 Myr, after which the accretion rate diminishes and growth plateaus. This stagnation coincides with the emergence of a thick sub-pc disc, visible in Fig. 4, which acts as a reservoir, accumulating gas that cannot be accreted on to the BH due to its angular momentum. In contrast, Halo 2 BHs exhibit a more gradual rise in accretion rate, peaking at 0.3 Myr. Despite periodic spikes, there is not a well-defined growth plateau. In fact, by the end of the period, a significant accretion event propels them to surpass their Halo 1 counterparts. A dense, $n \simeq 6 \times 10^7$ cm$^{-3}$ compact disc forms within the first 0.15 Myr in `2B.b01`, but is entirely absent in `2B.m01`. There is further discussion of the disc structure and properties in Section 3.2.

In both haloes, BHs employing the mass-flux model experience more efficient growth. Given that this scheme measures the mass falling within a shell twice the radius of $R_{BHL}$ ($r = 0.049$ pc and $r = 0.033$ pc for Haloes 1 and 2, respectively), this gas needs to shed less angular momentum to be captured by the BH compared to BHL accretion in simulations at marginal $R_{HL}$ resolution. Moreover, the BHL formula is more sensitive to the relative gas speed $\propto 1/v_\infty^3$. Early in the evolution of the BH, it will produce a lower accretion rate than the mass-flux scheme and more material will be left in the accretion region. Over time, this material will accumulate, gain angular momentum, and form a disc, further limiting accretion on the BH. This is the mechanism by which a disc is seen in 2B.b01 but not 2B.m01, and by which the disc is more dense in 1B.b01 than 1B.m01.

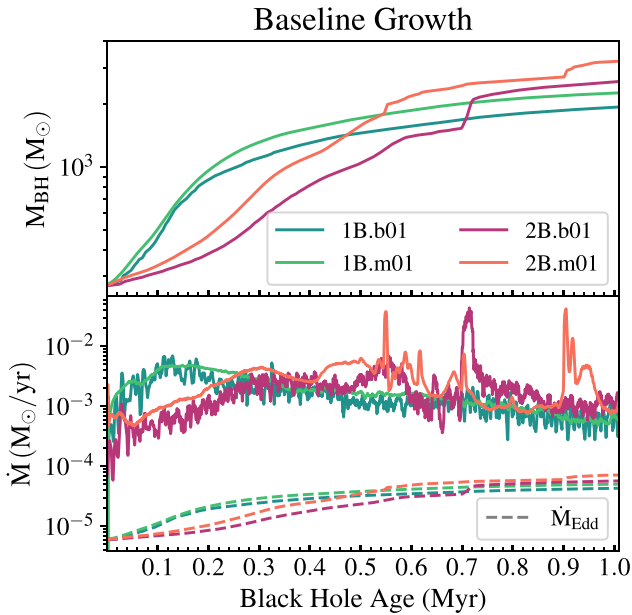

**Figure 3.** BH growth (panel 1) and BH accretion rate (panel 2) over 1 Myr of the 270 M$_\odot$ BH for the baseline resolution group of simulations, 1B.b01 (Halo 1, BHL accretion), 1B.m01 (Halo 1, mass-flux accretion), 2B.b01 (Halo 2, BHL accretion), and 2B.m01 (Halo 2, mass-flux accretion). The lower panel also includes the Eddington accretion rates as *dashed lines*. Halo 1 BHs experience a steeper initial growth period than Halo 2 BHs but plateau sooner.

*3.1.2 As resolution varies*

Fig. 5 shows the time evolution of the Halo 1 and Halo 2 270 M$_\odot$ BHs and their accretion regions when growing via BHL and mass-flux models. The simulation resolution is defined in units of Hoyle–Lyttleton ($R_{HL}$) radius resolution, where $R_{HL}$ is measured just before the BH particle is inserted e.g. 1B.x01 has the lowest HL radius resolution of just one cell and 1B.x16 has the highest, with $R_{HL}$ resolved by approximately 16 cells.

For both seeds and accretion schemes, growth in the first 0.5 Myr becomes less efficient as spatial resolution increases. While all Halo 1 BHs demonstrate a similar rapid accretion phase, the higher resolution simulations plateau sooner, which suggests more efficient nuclear disc formation and will be discussed further in Section 3.2. In Halo 2, the two highest and the two lowest resolution simulations exhibit very similar growth cycles, respectively. At $t \simeq 0.2$ Myr, the two groups diverge, with the accretion rate continuing to climb for the low-resolution runs, but dipping slightly for the high-resolution runs. At $t \simeq 0.5$ Myr, $2B.x08/16$ begins a period of elevated accretion with $\dot{M}_{BH} \simeq 1 \times 10^{-2}$ M$_\odot$ yr$^{-1}$. The gap between the low- and high-resolution groups reduces as a result. A similar phenomenon is observed in Halo 1, with $1B.x08$ and $1B.x16$ exhibiting step-like growth in the range $t \simeq 0.55–0.75$ Myr, causing these high-resolution runs to converge with $1B.x01$ by the end of the period. The only exception to this trend is 2B.b04 which grows more efficiently than 2B.b01 until 0.7 Myr when a bump in accretion rate in 2B.b01 causes it to overtake.

While the interplay between gas velocity and density differs slightly between the accretion schemes, with BHL simulations attaining higher densities and mass-flux attaining higher velocities, the resultant accretion rates are almost identical at high resolution. For instance, $1B.m16$ and $1B.b16$ both reach 1000 M$_\odot$ at 0.35 Myr, whereas $1B.m01$ reaches the same mass 0.4 Myr before $1B.b01$. The shrinking shell across which mass flux is measured reduces the accretion rate by up to a factor of 1/256–1/300 for 16-cell $R_{HL}$ resolution, while the density increases by a max of 100–200 times and the gas velocity only increases by a factor of a few times the lowest resolution values. The dominance of the shell radius size in determining the mass-flux accretion rate suppresses growth in the first 0.5 Myr. In the BHL simulations, the density increases up to a factor of 1000 from baseline resolution, while the Mach number maximally grows by a factor of 10 (the sound speed remains fairly constant at $c_{s,\infty} \simeq 1$ km s$^{-1}$). However, the sensitive dependence of BHL on the relative velocity, $\dot{M} \propto 1/v_\infty^3$, means that even this modest increase in $v_\infty$ considerably suppresses the accretion rate.

### 3.2 Black hole gas properties and disc morphology

*3.2.1 At baseline resolution*

A distinct disc structure can be identified early in the lifetime of the BH in 1B.b01 and 1B.m01, as illustrated in Fig. 4. The face- and edge-on density projections depict the evolution of the accretion region from a clump of overdense gas to a disc of density $\simeq 1 \times 10^7$ cm$^{-3}$ with radius $r_{disc} \simeq 0.2$ pc and thickness $h_{disc} \simeq 0.1$ pc after 1 Myr, which corresponds to the free-fall time of the gas within 1 pc of the centre of the halo before the BH particle is inserted. There are also signs of an outer disc of less dense gas, $n \simeq 4 \times 10^5$–$1 \times 10^6$ cm$^{-3}$, forming at this time, with a radius extending to $\simeq 0.4$ pc. From the final panel of Fig. 1, it can be seen that $r = 1$ pc marks the point at which the average density dips below $10^5$ cm$^{-3}$. The Halo 1 BHs remain nested in the centre of the cloud throughout their evolution.

In contrast, the Halo 2 BHs at baseline resolution (right two columns of Fig. 4) form a much smaller nuclear disc until over a longer period, with 2B.m01 barely forming a disc by the end of the 1 Myr. At $t = 0.3$ Myr, a dense, small disc-like structure is seen in 2B.b01, but is soon disrupted by tidal forces from a nearby clump. The elevations in accretion rate seen in panel 2 of Fig. 5(a) at 0.55 Myr for 2B.m01 and 0.7 Myr for 2B.b01 correspond with material ejected from this satellite being absorbed by the BH (as verified by a time series plot, not shown). Stability has somewhat recovered by $t = 0.8$ Myr in 2B.b01 and a clear disc is visible. In 2B.m01, dense gas has aligned with the orbital plane, and gradually over the next 0.3 Myr, it settles into a disc. While both Halo 2 discs are cylindrical and slim, with a disc radius $r_{disc} \simeq 0.2$ pc and height $h_{disc} < 0.1$ pc, they are not as symmetric nor uniform in density.

The primary morphological difference between mass-flux and BHL accretion in Halo 1 at baseline resolution is the underdense region surrounding the BH in 1B.m01 that is absent from 1B.b01. Indeed, from the second panels of Fig. 5, we see that the accretion region of 1B.m01 averages at $n \simeq 1 \times 10^7$ cm$^{-3}$, while 1B.b01 remains at $n \simeq 3 \times 10^7$ cm$^{-3}$. This is because gas is removed at each time-step from within $r = 4dx \simeq 0.05$ pc of the BH and reaches the 75 per cent limit on instantaneous mass removal each time (see Section 2.4). The continual evacuation of the accretion region leads to the formation of a 'hole' in the gas cloud around the BH of radius equal to the surface boundary used in the mass-flux scheme. This is not a physically realistic phenomenon as such a large region should not be accreted by a BH of this mass at such regular intervals; it is simply an artefact of the model. The BHL scheme does not have this issue. From panel 2 of Fig. 5(a), we see that the BHL accretion rate can vary by up to an order of magnitude within 10 kyr. Likewise, the gas density and velocity are more variable than their mass-flux counterparts. The sensitivity of BHL to gas velocity means that, as gas is removed from the grid and added to the BH, the sudden decrease in pressure causes an increase in velocity, which in turn lowers the accretion rate. As gas accumulates again, the velocity drops and density rises, resulting in efficient accretion once more. The BHL model allows the gas in the accretion region to be replenished periodically and only a modest reduction in density is observed in 1B.b01 (see Fig. 6, upper right panel).

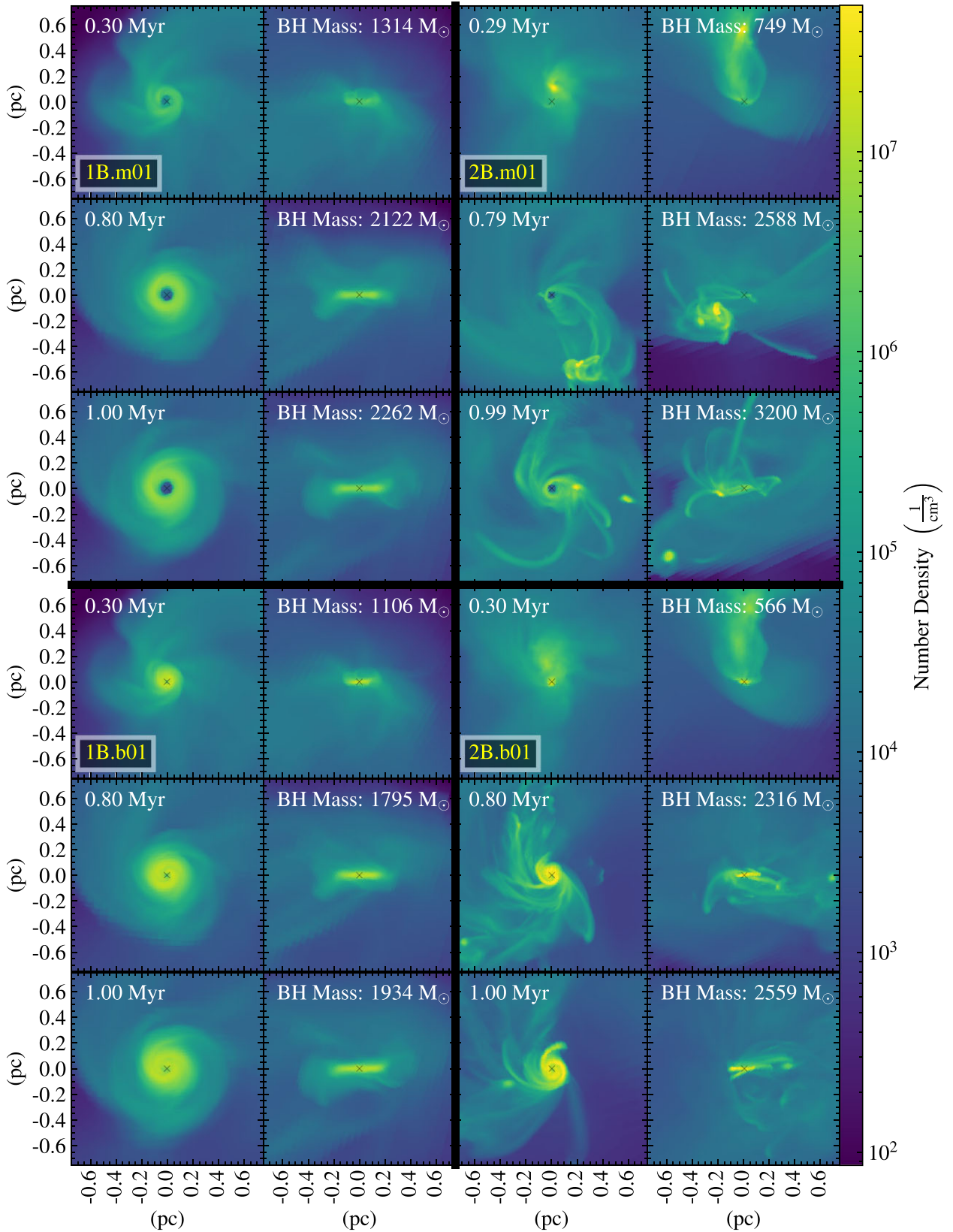

**Figure 4.** Total hydrogen nuclei number density projections face-on (first and third column) and edge-on (second and fourth column) of the collapsing cloud in `1B.m01` (top left two columns), `1B.b01` (lower left two columns), `2B.m01` (top right two columns), and `2B.b01` (lower right two columns). Time increases from top to bottom in variable increments, as shown in the first and third columns for each simulation. The BH location is marked by a *black cross* in each panel and its growing mass is indicated in the first and third columns.

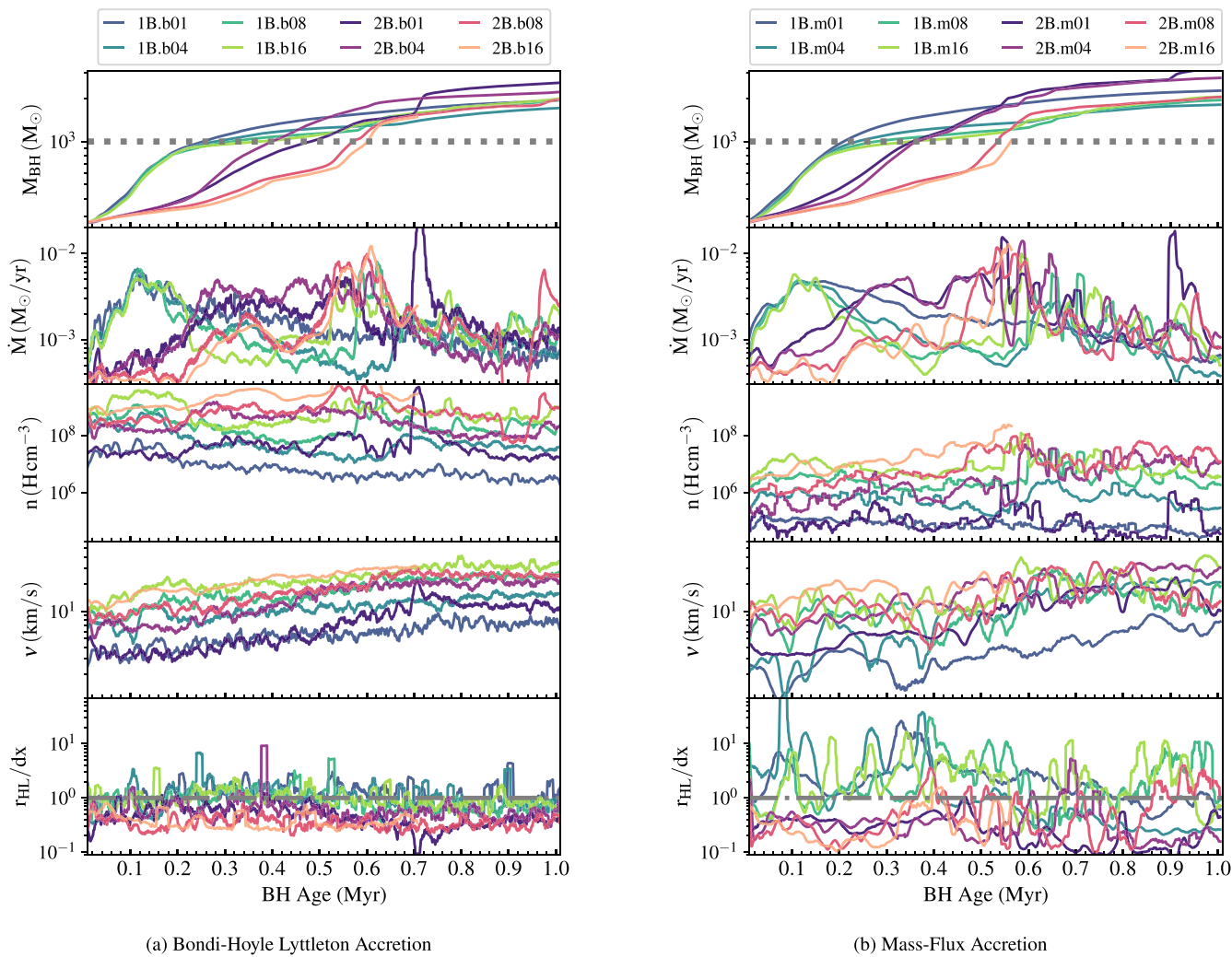

(a) Bondi-Hoyle Lyttleton Accretion

(b) Mass-Flux Accretion

**Figure 5.** Time evolution of mass-averaged quantities in the accretion region of the 270 $M_\odot$ BH growing via the BHL scheme (left) and the mass-flux scheme (right). The four simulations in each sub-plot are identical except for their spatial resolution; 1B.x01 has the lowest, 1B.x16 the highest, with a factor of roughly 16 difference between their cell widths (see Table 4 for further details). From top to bottom, the panels illustrate the mass of the BH, the accretion rate of the BH, the hydrogen nuclei number density of the gas in the accretion region, the velocity of the gas with respect to the BH, and the Hoyle–Lyttleton radius resolution $r_{HL}/dx$, where $dx$ is the cell width of accretion region. The second panel includes the Eddington rate $\dot{M}_{Edd}$ as a *dashed line*. The data have been time-averaged and interpolated so that an equal number of points (∼4000) are plotted for each line.

In the Halo 2 simulations, we find analogous differences between the mass-flux and BHL modes of accretion as observed in Halo 1, but with the additional observation that BHL leads to more efficient disc formation than mass-flux. A 'hole' can be identified at $t = 1$ Myr for 2B.m01 in the upper right of Fig. 4. Its BHL counterpart (lower right of Fig. 4) forms a much denser $n > 1 \times 10^7$ cm$^{-3}$ disc structure with no apparent underdense region at the centre. The evolution of gas properties in Fig. 5 for the Halo 2 baselines follow the trends of their Halo 1 counterparts; 2B.m01 remains at densities $n < 10^5$ cm$^{-3}$, while 2B.b01 oscillates just above $n \simeq 10^7$ cm$^{-3}$. When clumps of gas are absorbed by the BH at $t = 0.7$ Myr in 2B.b01 and $t = 0.55$ Myr in 2B.m01, the number density spikes temporarily by about two orders of magnitude, reflecting the density of the clumps. The less efficient accretion of the 2B.b01 BH in comparison to 2B.m01 may be linked to the dense accumulation of gas around it from an early stage. Due to gravitational interaction with a nearby clump of gas between $t = 0.4$ Myr, smooth evolution into a $r \sim 0.1$ pc disc (as in the Halo 1 simulations) is prevented. However, a new gas supply from a direct collision with the companion clump in the range $t = 0.7$–$0.8$ Myr leads to a spike in BH mass (see Fig. 3) and the extension of the disc. In contrast, `2B.m01` accretes more during the interaction with the gas clump and only collides with it at $t \simeq 1$ Myr, at which point we see a disc forming for the first time in its evolution.

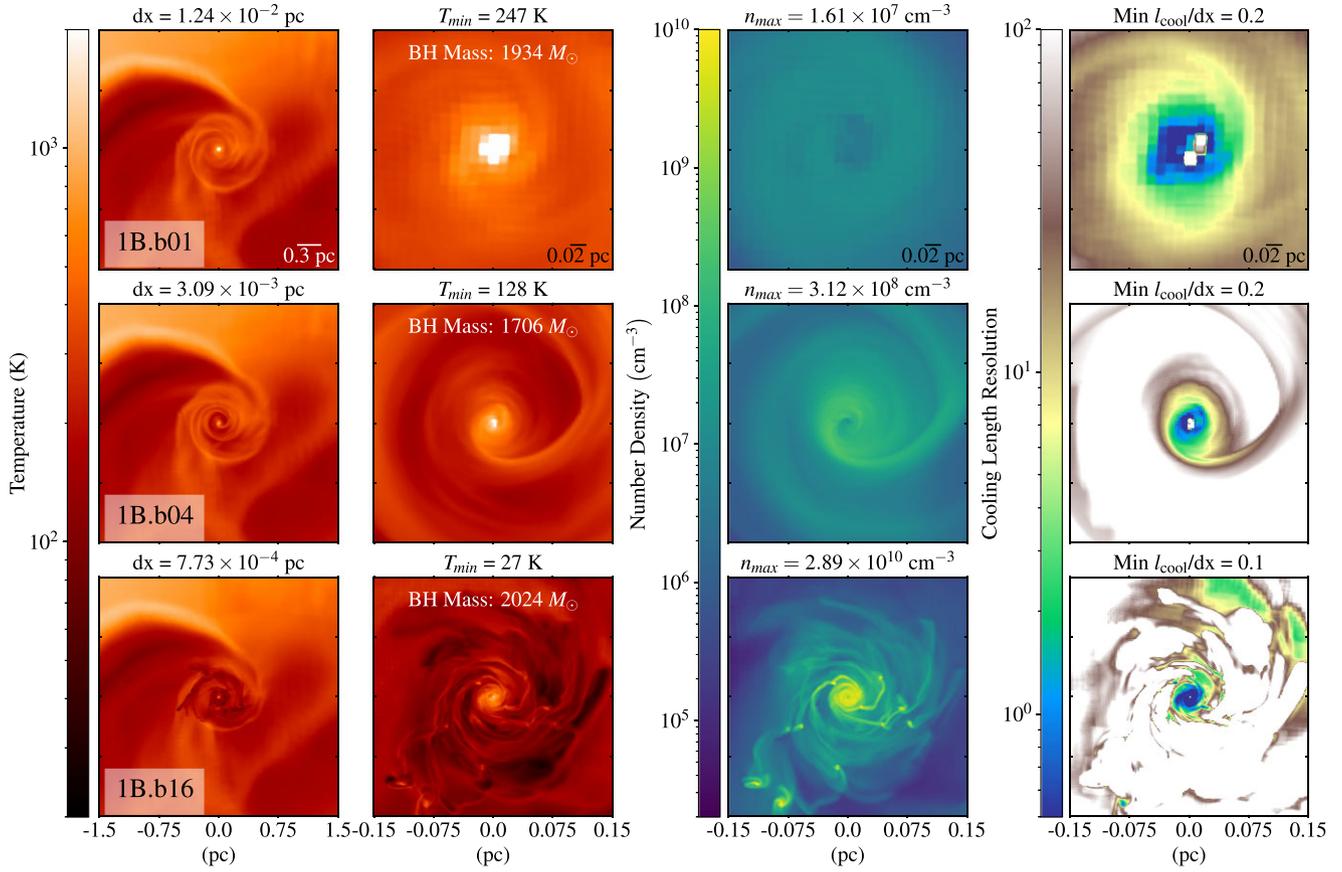

**Figure 6.** Projections centred on the BH with increasing simulation resolution (from top `1B.b01` to bottom row `1B.b16`) at $t = 1\,$Myr. Each column is at a certain level of zoom-in, as indicated by the scalebars and axes labels. The first two columns show the temperature, the next the number density, and the last the cooling length resolution. The column titles are the cell width of the most refined grid, minimum temperature reached, BH mass, and the minimum cooling length resolution. Regions of cold ($T < 100$ K) gas can be identified in the highest resolution simulation that are not seen in its lower resolution counterparts. Likewise, `1B.b16` attains densities exceeding $10^{10}\,$cm$^{-3}$ around the BH and in discrete clumps in the nuclear disc.

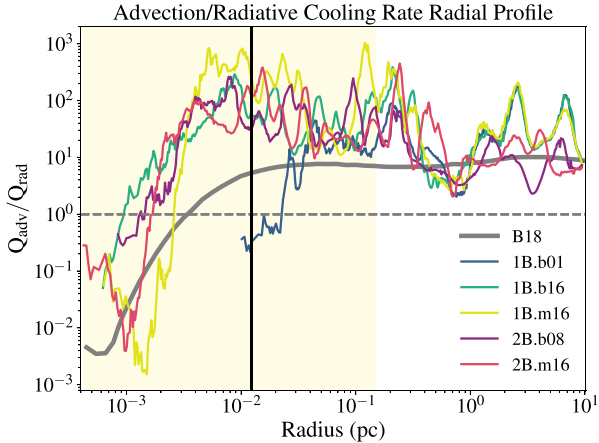

**Figure 7.** The relative dominance of advective $Q_{\rm adv}$ over radiative $Q_{\rm rad}$ cooling throughout the nuclear disc of `1B.b01`, `1B.b16`, `1B.m16`, and `2B.b08` at BH Age = 1 Myr. The *grey dashed line* corresponds to $Q_{\rm adv}/Q_{\rm rad} = 1$, below which radiative cooling dominates. The *black vertical line* indicates the extent of the accretion region for this 270 M$_\odot$ BH at $r = 0.012$ pc. The *grey solid line* is taken from fig. 15 in B18b. It was produced from fitting data in a simulation of resolution $r = 0.015$ pc, hence, beyond this point (roughly the black vertical line), the fit is extrapolated to unresolved scales and constitutes a prediction. The *yellow shaded region* illustrates the approximate size of the disc in `1B.b01`.

### 3.2.2 As resolution varies

As simulation resolution increases, more clumps of gas form in the vicinity of the BH. This is true for BHs in both haloes and accretion schemes. As discussed in the previous subsection, even at its baseline resolution of $dx = 8.3 \times 10^{-3}$ pc, Halo 2 simulations form some clumps, whereas the Halo 1 baselines at $dx = 1.23 \times 10^{-3}$ pc do not, indicating a characteristic resolution for clump formation of $dx \simeq 8 \times 10^{-3}$ pc. Fig. 6 depicts the variations in temperature, number density, and cooling length resolution $l_{\rm cool}/dx$ for the `1B.b` set of simulations at $t = 1$ Myr. The resolution increases by a factor of 4 per row, with the first row at the Halo 1 baseline and the last at the maximum resolution. Once we start zooming in from pc to sub-pc scales, we see that `1B.b16` has a much clumpier disc structure than even the intermediate `1B.b04`. These latter two simulations reach much higher densities than 1B.b01, with $n \geq 10^9$ cm$^{-3}$ in the arms of the disc. When these clumps fall into the BH, there are spikes in the accretion rate. In terms of BH growth, the highest resolution run is the most massive by the end of the simulated period, but only just. The extent of the nuclear disc, as estimated from the middle column on $\sim 1$ pc scales, is more or less consistent as resolution increases, with $r_{\rm disc} \simeq 0.5$ pc.

The manifestation of more spiral arms and gas clumps in the disc as resolution increases correlates with a decrease in temperature and an increase in the cooling length resolution. In the second column of Fig. 6, we see that the region beyond $r \simeq 0.5$ pc looks very similar at each resolution, but the disc itself differs notably (this is not surprising as the high-resolution region probably does not reach beyond the disc, see Fig. 2). The 1B.b16 simulation has extended regions of very cold gas ($T < 100$ K), while 1B.b04 and 1B.b01 remain within $T = 100$–1000 K (though the central 0.4 pc in 1B.b01 becomes heated to $T > 2000$ K). The final column reveals that the cooling length, $l_{\rm cool} = c_s t_{\rm cool}$, is at least marginally resolved throughout most of the disc at each resolution. However, in the region $0.02 < r < 0.075$ pc, better resolving the cooling length from $l_{\rm cool}/{\rm d}x \sim 1$ in the top row to $\sim 10$–100 in the second and third rows has induced the formation of instabilities. Therefore, it is crucial to resolve the cooling length by a factor of a few to accurately capture the multiphase nature of the ISM. Notably, the inner $r < 0.02$ pc region around the BH remains underresolved at all spatial resolutions, with the minimum cooling length resolution remaining at $l_{\rm cool}/{\rm d}x \simeq 0.2$ (as shown above each panel in column 4). When the simulation resolution increases, the gas condenses to a lower temperature and higher density than before, driving down the cooling length (see fig. 8 of Smith et al. 2017). When ${\rm d}x > l_{\rm cool}$, gas fragmentation due to cooling is smoothed, whereas when ${\rm d}x < l_{\rm cool}$, instabilities can proliferate. Indeed, from the top-left panel of fig. 8 in Smith, Sigurdsson & Abel (2008), we can glean that the classical fragmentation criterion, $t_{\rm cool} < t_{\rm dyn}$, is well satisfied for 1B.b16, which has regions where $T < 100$ K and $n > 10^8$ cm$^{-3}$ and not satisfied for 1B.b01, which has much hotter core and $n \leq 10^7$ cm$^{-3}$. The consistent minimum $l_{\rm cool}/{\rm d}x$ reached in these simulations indicates that the gas self-stabilises at a certain point due to a maximum refinement level being imposed. Should this be relaxed, the gas would continue to cool and fragment. The highest resolution simulation marginally underresolves it close to the BH, where the gas is very dense and hot. This ensures that we are largely capturing the multiphase ISM, as failing to do so may lead to misrepresentations of the gas flow and clumping (Rey et al. 2024). This considerable variation in temperature with resolution is due to the increased efficiency of the three-body channel of $H_2$ formation at high densities.

Despite the differences in the gas attributes as resolution increases, both accretion schemes reach a similar final mass at high resolution to each other and their respective lower resolution runs. As we have seen, the mass-flux scheme continually evacuates the accretion region, leading to a lower average density in the accretion region (see the underdense 'hole' in Fig. 2 and panel 3 of Fig. 5b). While the schemes have the same $\rho$ dependence, they have inverse relations to velocity: $\dot{M}_{\rm flux} \propto v^-$, whereas $\dot{M}_{\rm BHL} \propto 1/v^3$. The gas is fast moving in both the circular and radial components, which will cause $\dot{M}_{\rm flux}$ to be large and $\dot{M}_{\rm BHL}$ to be comparatively small in *the first time-step*, when both are accreting from the same environment. Indeed, we can see clearly that 2B.m01 has a much higher accretion rate than 2B.b01 in the first few years in Fig. 3. Hence, more gas will be removed by the mass-flux model in that time-step than for BHL. In the subsequent time-step, the gas density in the accretion region of the mass-flux simulation will be comparatively lower, which will limit $\dot{M}_{\rm flux}$. While the BHL simulation now has a more dense accretion region, $\dot{M}_{\rm BHL}$ is still suppressed. As gas is removed from the grid, the decrease in pressure causes an increase in $v$ which can considerably reduce $\dot{M}_{\rm BHL}$, given its $\propto 1/v^3$ dependence. The greater variability of $\dot{M}_{\rm BHL}$ in comparison to $\dot{M}_{\rm flux}$ is likely due to this sensitive dependence on gas velocity. These two distinct mechanisms of self-regulation result in a similar overall accretion pattern between the schemes at each resolution level. The reason $\dot{M}_{\rm BHL}$ and $\dot{M}_{\rm flux}$ converge as resolution increases is related to the shrinking accretion radius. The mass flux rate continuously removes the maximum allowed gas from within the accretion region (tracked explicitly in the simulation) and is thus supply-limited. With a smaller physical region to accrete from at high resolution, $\dot{M}_{\rm flux}$ becomes suppressed. Gas must lose more angular momentum in the disc to reach the smaller spherical boundary. The BHL rate is likewise suppressed due to the higher peak in gas velocity as scales closer to the bottom of the potential well are probed at high resolution. This limited accretion rate at the start of the evolution is counteracted by 'episodes' of accretion later on when the many clumps that form at high resolution lose enough momentum to fall in.

Fig. 7 shows the ratio of advective to radiative cooling as a function of radial distance from the BH for Halo 1 and Halo 2 simulations at the end of their evolution. The radiative cooling rate $Q_{\rm rad}$ represents the total energy lost per unit volume per unit time due to the emission of photons, and is calculated directly by the chemistry solver. The advective cooling rate $Q_{\rm adv}$ measures the amount of thermal energy carried away per unit time by the bulk fluid motion and is calculated according to the formula

$$Q_{\rm adv} = \frac{\rho v_r T k_{\rm B} \xi}{m_p r} \quad [{\rm erg\,s^{-1}\,cm^{-3}}], \tag{14}$$

where $\rho$, T, $v_r$, and r are the mass density, temperature, radial velocity, and radius from the BH, $\xi = -0.65$ is a dimensionless parameter following Chen et al. (1995), $k_{\rm B}$ is the Boltzmann parameter, and $m_p$ is the mass of a proton. In Fig. 7, the *grey solid line* is taken from fig. 15 of B18b. A key difference between our chemistry and that of B18b is our use of GRACKLE, which solves for the cooling rate directly from the chemistry network and includes the three-body mechanism of $H_2$ formation. A density cap of $10^9$ cm$^{-3}$ was imposed on all B18b simulations to reflect the omission of this reaction (which gains significance beyond this density). Nevertheless, they used estimations of the expected increase in $f_{H_2}$ at higher densities to extrapolate their fit down to scales below their resolution (${\rm d}x = 0.015$ pc, almost identical resolution to baseline Halo 1), as shown by the *solid black line* in Fig. 7. In keeping with this predicted trend, our simulations stay largely advection-dominated within the disc (*yellow region*). However, radiative cooling begins to grow in prominence approaching the resolution limit in all simulations. Our high-resolution simulations (all bar 1B.b01), in which densities can exceed $\sim 10^9$ cm$^{-3}$, experience radiative cooling domination on scales within radii two to three times closer to the BH than the B18b fit. The discrepancies between our data and the fit could be due to environmental differences between our simulation suite and that of B18b. They use SMBH scale BHs ($\sim 10^6$ M$_\odot$) and haloes ($\sim 10^{11}$ M$_\odot$) to produce the trendline; the stronger gravitational field combined with the greater gas supply may lead to a wider extent of cold, dense, radiative-cooling-dominated gas around the BH.

While all Halo 1 simulations form discs on roughly the same time-scale, Halo 2 simulations accreting via mass-flux form discs more efficiently as resolution increases. As noted in Section 3.2.1, a distinct disc structure cannot be identified in 2B.m01 until almost $t = 1$ Myr. At a resolution four times greater in 2B.m04, a dense accumulation about $\simeq 0.05$ pc in extent forms much earlier at $t = 0.3$ Myr. From the final panel of Fig. 8(b), we observe the increasing radiative cooling rate from large to small radii until approximately quadruple the spatial resolution of each simulation is reached, which corresponds to the radius of the accretion region; 2B.m01 (*purple line*) declines sharply in cooling rate at $r = 0.04$ pc while 2B.m04 does not drop off until $r = 6 \times 10^{-3}$ pc. Thus, as resolution increases, the underdense 'hole' around the BH grows smaller and matter needs to travel further to be absorbed by the BH. As gas moves inwards, it gains angular momentum, reaching a circular velocity Mach number of 20 (see panel 4 in Fig. 8b), and begins to accumulate instead of falling directly into the accretion region. The more efficient radiative cooling allows the gas to reach temperatures $T < 1000$ K and densities $n \simeq 1 \times 10^8$ cm$^{-3}$, helping the gas lose enough energy to settle into a stable orbit.

### 3.3 Disc instabilities

Early in the evolution of 1B.m08/16 and 1B.b08/16, the disc resembles that of the lower resolution 1B.b01 as seen in the time series of Fig. 4; almost uniform in density, thick, circular, and featureless. In high-resolution simulations, as the density increases, the radiative cooling rate also increases (see panels 2 and 7 of Fig. 8). This leads to the development of gas instabilities in the form of dense spiral arms and bar-like structures which help redistribute angular momentum from the centre to the outer parts of the disc. While the lower resolution simulations also display spiral arms in their discs (see Figs 4 and 6), there is little fragmentation and no bar-like structures. More precisely, no bar formation was observed at a resolution lower than $dx = 1.54 \times 10^{-3}$ pc in either halo.

Bars and spiral arms are usually discussed in the context of galaxy simulations on $\sim$ kpc scales (Petersen, Katz & Weinberg 2016) or protoplanetary discs on $\sim$ AU scales (Saigo, Hanawa & Matsumoto 2002); however, we can adapt certain metrics for our pc-scale discs. In a similar manner to Athanassoula & Misiriotis (2002) and Ansar et al. (2023), we identify the presence of these features as elevations in the $m = 2$ Fourier mode amplitude of the 2D decomposition of the surface density of the disc, though with the surface composed of gas cells rather than stellar particles. The Fourier mode amplitudes are given by

$$\frac{A_m}{A_0} = \frac{\sqrt{a_m^2 + b_m^2}}{\sum_{i=1}^{N} m_i}, \quad (15)$$

where, in general for the $m$th mode,

$$a_m = \sum_{i=1}^{N} m_i \cos(m\theta_i),$$

$$b_m = \sum_{i=1}^{N} m_i \sin(m\theta_i), \quad (16)$$

where $\theta_i$ is the azimuthal angle and $m_i$ is the mass of the $i$th cell. The $m = 1$ mode represents a one-fold symmetry, often corresponding to an off-centre displacement or one-sidedness in the system. The $m = 2$ mode represents a two-fold symmetry, indicating the presence of a bar or dual spiral arm structure.

We attempt to distinguish between bars and spirals as follows. We require that a bar-like feature must have an average $A_2/A_0 \geq 0.70$ within $r = R_{\rm bar}$, where $R_{\rm bar}$ is defined as the radius of an overdensity with a near-constant $m = 2$ phase angle $\phi_2$ extending to at least $r = 0.01$ pc from the BH particle. The bar radius is found by calculating $\phi_2 = \arctan(b_2/a_2)/2$ within narrow annuli of increasing radius from the centre and then identifying the point at which the cumulative standard deviation exceeds 7°. Fig. 9 shows a time series of 1B.m16 simulation snapshots, with each row corresponding to the time indicated in the third column. The first two columns illustrate how $R_{\rm bar}$ and $R_{\rm co-rot}$ are found, respectively, with $R_{\rm bar}$ calculated from the $m = 2$ Fourier mode as just outlined. The co-rotation radius $R_{\rm co-rot}$ is defined as the intersection of the angular frequency of the disc and the mean bar pattern frequency, and is found by truncating the $m = 2$ pattern frequency radial profile at $r = R_{\rm bar}$ and taking the average. This is represented by the *red dashed line* in column 2 of Fig. 9. We also confirm that the bar pattern frequency is lower than the angular frequency of the disc with a co-rotation radius $5R_{\rm bar} < R_{\rm co-rot} < 25R_{\rm bar}$, which is consistent with the expected dynamics of a slow-moving bar that does not dominate the local potential. Spiral arms are defined both spatially and temporally; they manifest in regions of the disc where the $m = 2$ mode dominates over the $m = 1$ mode (hence we are excluding one-arm spirals) and must have an initial perturbation amplitude of $A_2/A_0 \geq 0.5$, which can fall over time as it moves outwards and still be considered a dual arm. These features are verified visually by examining density slices through the plane of the BH in column 3 of Fig. 9. Column 4 shows the value of the Toomre Q instability parameter, which is defined as

$$Q = \frac{c_s \kappa}{\pi G \Sigma}, \quad (17)$$

where $c_s$ is the sound speed, $\kappa$ is the radial frequency, and $\Sigma$ is the surface density. If $Q \leq 1$, the disc is unstable to axisymmetric perturbations, so small density enhancements can grow under their own gravity to form spiral arms, bars, or clumps. In general, when $Q > 1$, the combination of thermal pressure and differential rotation acts to counterbalance gravitational instabilities. However, in the range $1 < Q < 10$, the disc is still susceptible to the formation of non-axisymmetric small-scale instabilities if the Mach number is sufficiently high (Hopkins & Christiansen 2013). Indeed, the spiral arms at BH Age = 0.5 Myr are traced by values within this range, with the lowest values occurring in the dense inner regions and bar structure. At BH Age = 0.59 Myr, the clumps in the disrupted disc fall to $Q < 1$ and are therefore unstable; some are destined to self-gravitate and remain within the disc (see column 3 of Fig. 6), while others are flung out. The bar instability remains traced by low Toomre Q values at later times, though becomes thinner and curled at BH Age = 0.96 Myr, when, by our criteria, the bar structure is depleting. The final column shows the radial velocity distribution, where he negative values represent gas moving towards the BH. The gas along the bar is, in general, slow-moving and slightly positive, whereas the gas on either side of the bar is strongly in-falling with $v_{\rm r} \geq -10$ km s$^{-1}$.

Fig. 10 summarises the frequency and duration of bar-shaped and spiral arm features and disc fragmentation and stability in 1B.m16. We will now analyse the evolution of these instabilities and their impact on BH growth. As previously discussed, the first $t \simeq 0.15$ Myr of BH evolution in 1B.m16 is characterised by increasingly rapid growth during which most of the 600 M$_\odot$ core is accreted. The gas has a high radial velocity component, falling on to the BH in a spherically symmetric manner. As indicated with the *pink region* in Fig. 10, gas accumulates along a certain axis to form a thick disc structure at $t \simeq 0.18$ Myr, which is determined by inspecting face- and edge-on density projections as in Fig. 4. While the disc formation process is associated with a declining accretion rate, the onset of spiral density waves from the centre of the disc at $t = 0.32$ Myr corresponds with a marked uptick. The development of a bar-shaped instability across the accretion region coincides with a spike in accretion rate between

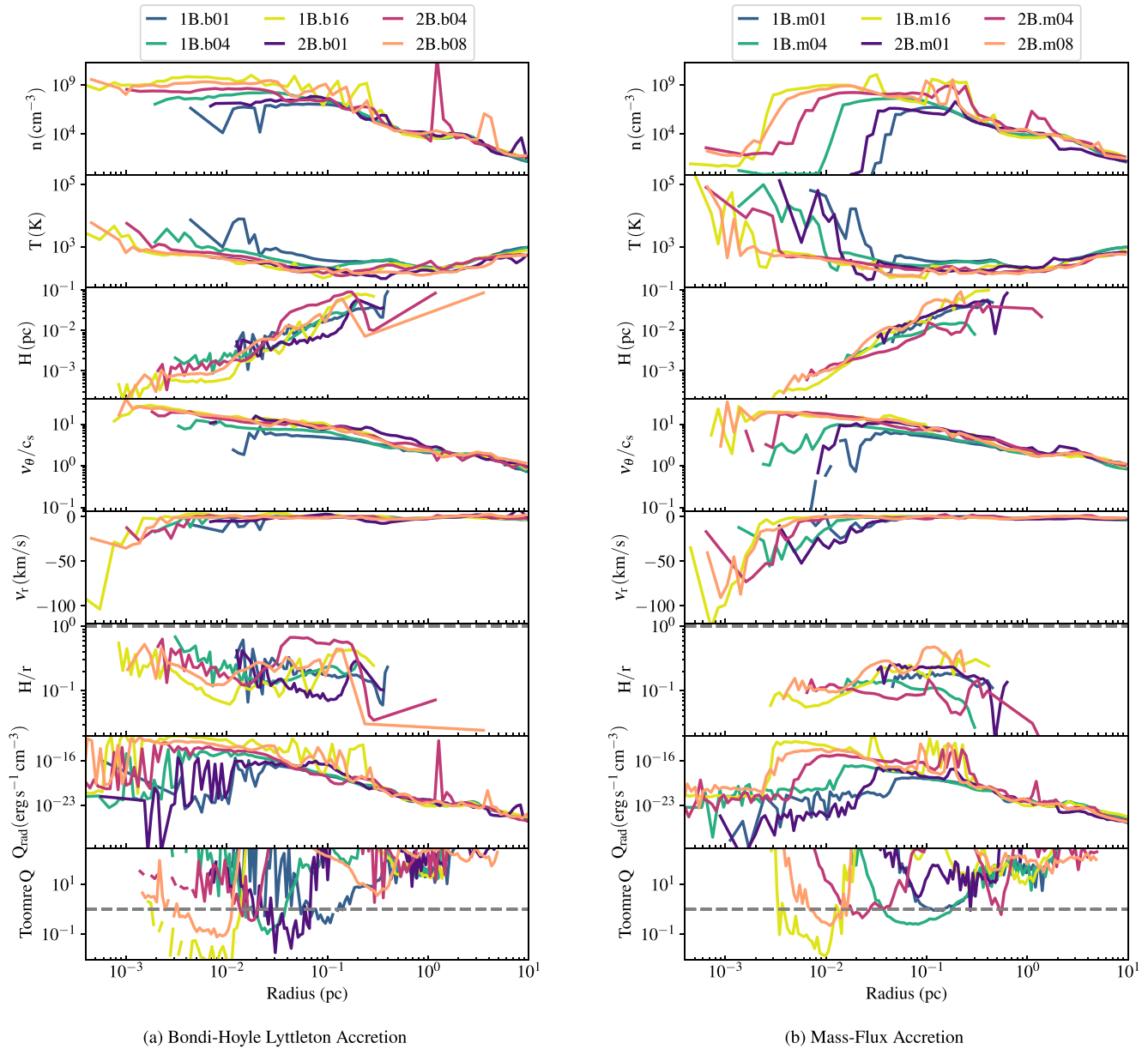

**Figure 8.** Radial profile of disc attributes at BH Age = 1 Myr in Halo 1 and Halo 2 BHs accreting via BHL and mass-flux at various levels of refinement. From top to bottom, the panels depict hydrogen nuclei number density, temperature, disc height, the ratio of circular velocity to sound speed, radial velocity, the ratio of disc height to radius, and the radiative cooling rate. The disc height is measured from a cylindrical region centred on the BH in which the condition $n > n_{max}/100$ is imposed, where $n_{max}$ is measured from within $r = 10\,dx$.

0.48 and 0.52 Myr. The first row of panels in Fig. 9 shows the state of the disc in the middle of this period. According to the criteria explained in the previous paragraph, this instability is considered bar-like as its radius exceeds 0.01 pc, the amplitude of the $m = 2$ Fourier mode exceeds 0.8, and a distinct rectangular feature can be seen in the density slice. The Toomre Q shows the gas along the bar and the spiral arms to be marginally stable, while the radial velocity map clearly depicts the rapidly infalling gas on either side of the bar being funnelled towards the BH. The bar can therefore originate instabilities to redistribute angular momentum away from the BH, promoting accretion on to it. This inward torquing of gas within the co-rotation radius has been shown to occur in simulations of bars formed via galaxy mergers (Hopkins et al. 2009) and in studies of idealised bars (e.g. Berentzen et al. 2007). In less than

0.09 Myr, however, fragmentation spreads throughout the entire disc, with complex interactions occurring between clumps; some are flung out, some fall inwards and get absorbed by the BH and others become unbound and smoothed into the disc. The interactions become intense enough to temporarily destroy the bar feature and disrupt the stability and symmetric morphology of the disc. This results in a more oscillatory accretion rate from $t \simeq 0.52$–0.78 Myr and episodic growth cycle (see first panel of Fig. 5b for the 'bumpy' growth during this period). The bar and dual spiral structures recover periodically in the latter half of the evolution. The third row of Fig. 9 shows a point at which a central instability just qualifies as a bar feature with radius $R_{bar} = 0.016$ pc. The spiral arms also become more tightly wound, almost forming rings, which can be seen as peaks in the $m = 2$ Fourier mode on the leftmost plot. By 0.96 Myr

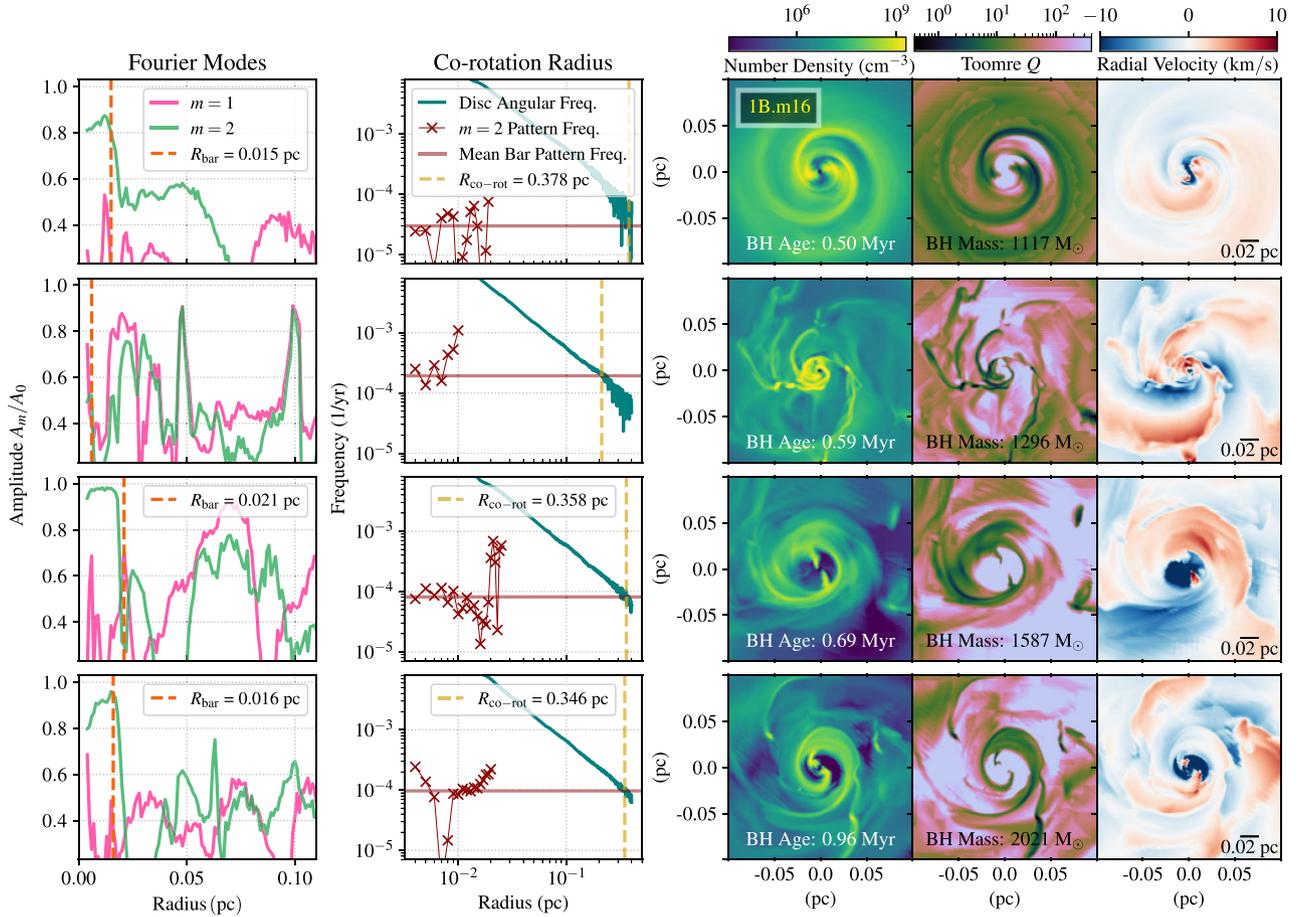

**Figure 9.** Time series of slices (row-wise) through the plane of a $r = 0.1$ pc disc in `1B.m16` centred on the BH particle. The first two columns are radial profiles; the leftmost shows the normalised $m = 1, 2$ Fourier mode amplitudes used to determine the bar radius $R_{\rm bar}$ (as discussed in Section 3.3) indicated with an *orange dashed line*, while the second column plots the disc rotational frequency and the bar pattern frequency, the intersection of which defines the co-rotation radius $R_{\rm co-rot}$ shown as a *yellow dashed line*. The mean bar pattern frequency is calculated by truncating the $m = 2$ list of phase angles measured per annulus at $r = R_{\rm bar}$ and taking the average value. The remaining three columns are maps of total hydrogen nuclei number density, the Toomre Q stability parameter, and radial velocity, respectively. A dense, bar-shaped instability forms after $t \simeq 0.5$ Myr. While the feature is disrupted during a period of intense fragmentation around $t \simeq 0.54$ Myr, it recovers intermittently during the last 0.35 Myr, with strongly wound spirals emanating from its poles.

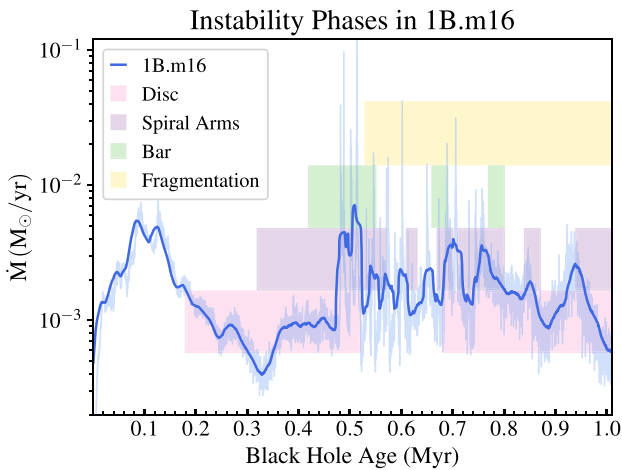

**Figure 10.** Presence and duration of the disc and various kinds of gas instabilities in `1B.m16` overlaid on the BH accretion rate versus time.

in the final row, the bar feature has twisted more at its poles, forming an elongated s-shape.

Bar-shaped instabilities are also seen in the Halo 2 simulations `2B.b08/16` and `2B.m08/16`, though they are smaller and less enduring than in Halo 1. The primary difference in the evolution of Halo 2 at high resolution is the collapse of a clump of gas offset from the BH position. A compact, axis-symmetric disc less than 0.01 pc in extent forms around the BH, quickly reaching high densities. Around the same time as in Halo 1, BH Age $\simeq 0.3$ Myr, a bar-shaped structure and spiral arms are first seen, and thereafter, the disc loses its stable structure. By 0.6 Myr the compact central disc is recovered, though with no apparent instabilities. In contrast to Halo 1, over the next 0.2 Myr, a second disc forms around it at a polar angle (i.e. on a different plane of rotation) with gas siphoned from the nearby overdensity. A recent analysis of a TNG catalogue of galaxies indicates that polar rings form from 'close interactions with a gas-rich companion or satellite', which is what we have observed on much smaller scales in Halo 2 (Smirnov, Mosenkov & Reshetnikov 2023). From 0.89 to 0.95 Myr, the accretion rate falls in both `2B.b08` and `2B.m08` (see second panels of Fig. 5), which aligns with the integration of the inner compact disc into the outer disc's plane. Between 0.92 and 0.95 Myr, a $R_{bar} < 0.01$ pc s-shaped bar overdensity is seen in slices through the disc, similar to the one at the end of the `1B.m16` evolution, though not as prominent nor long. Overall, the evolution of instabilities in Halo 2 resembles that of Halo 1 in the first 0.5 Myr, though the non-secular formation of a polar ring and infalling gas clumps results in some distinct differences in the latter half of the period.

### 3.4 Post-Type II supernova black hole

We now compare our findings for the 270 $M_\odot$ BH with that of a 10.8 $M_\odot$ BH. This smaller black hole is the remnant of a 40 $M_\odot$ Pop III star undergoing core-collapse at the end of its main-sequence lifetime (Woosley, Heger & Weaver 2002). The companion Type II supernova releases a few solar masses of metals into the surroundings, enriching the pristine gas cloud. Since the main-sequence lifetime of the progenitor star was not simulated, the supernova occurs in a gas-dense environment which limits the propagation of the Sedov–Taylor blast wave ($R \propto 1/\rho^5$) (Hayes et al. 1967). As a result, gas heating is minimal. This section investigates the short-term growth of the resulting 10.8 $M_\odot$ BH both with and without a supernova event preceding it to isolate the effects of the burst of thermal feedback and metal enrichment of the gas. The relevant simulations are detailed in Table 5.

#### 3.4.1 At baseline resolution

At baseline resolution, comparatively modest growth is observed for the 10.8 $M_\odot$ BHs with SN feedback in both sets of initial conditions and accretion schemes, as per Fig. 11. In Halo 1, `1S.m01` grows more efficiently than its BHL counterpart, as we saw for the 270 $M_\odot$ BH, reaching twice the mass by t = 1 Myr. When there is no supernova event nor metal enrichment (i.e. the simulations with the `-no-SN` suffix), the BH growth is similarly modest and follows a similar accretion cycle. However, the `1S.m01-no-SN` BH grows to almost double the mass of the `1S.m01` BH. We can see from Fig. 12 that the gas cloud in `1S.m01-no-SN` condenses more slowly than in `1S.m01`, so the BH is 'in contact' with dense gas for longer and can accumulate more mass from the clump before it moves away. In contrast, the Halo 2 `-no-SN` BHs grow less efficiently than their counterparts with SN feedback. The largest discrepancy occurs between `2S.b01` and `2S.b01-no-SN`, where the former grows to over four times the mass of the latter. The metal cooling and SN feedback which led to a reduction in the gas supply to the Halo 1 BHs helped promote accretion on to the Halo 2 BHs. Clearly, there is no consistent growth trend when SN feedback and metals are added; the growth of each BH is sensitive to small-scale dynamics.

#### 3.4.2 As resolution varies

As resolution increases in Halo 1, a greater final mass is reached by both accretion schemes. In Fig. 13, we can see that `1S.b04` and `1S.m04` reach the dotted horizontal line at $M_{BH} = 60 M_\odot$ by the end of the period, though `1S.m04` reaches it about 0.1 Myr earlier. This convergence underscores the trend observed with the 270 $M_\odot$ BH; at high resolution, the particular accretion model used becomes less important, as long as its input gas attributes are measured locally. The `1S.m04` simulation has a less efficient initial growth phase than `1S.m01`, but overtakes during an accretion period from 0.55–0.80 Myr caused by gas from the clump fragmenting and being ejected towards the particle. This dispersion of gas around the particle makes it more likely to grow than the lower resolution simulations in which the gas clump remains bound and drifts away. The Halo 1 BHs without supernova feedback, in contrast, do not converge in mass across accretion schemes at higher resolution; `1S.m04-no-SN` grows to $\simeq 10^3 M_\odot$ while `1S.b04-no-SN` barely reaches 20 $M_\odot$. The former is the only simulation out of all marginal to well-resolved $R_{HL}$ runs that gravitationally captures and accretes most of the dense cloud core as the 270 $M_\odot$ particle does consistently within 1 Myr. The degree of fragmentation of the gas leads to a direct collision with a large clump at t = 0.26 Myr, triggering a leap in mass and the formation of a disc structure. With increasing resolution comes a greater degree of fragmentation, more complicated interactions between clumps, and therefore a more varied and unpredictable BH accretion history.

As resolution decreases in both haloes, the growth patterns diverge between the accretion schemes. The `1S.bf4` and `1S.bf8` BHs barely increase in mass as their accretion rate quickly becomes sub-Eddington. In contrast, `1S.mf4` and `1S.mf8` experience a 0.2 Myr rapid growth period and accrete most of the $\simeq 10^3 M_\odot$ core. As previously alluded to, this is due to the mass-flux being measured across an increasingly wide sphere far from the BH, overestimating the quantity of mass flowing towards the BH [$M_{flux} \propto v^-(r)$]. The BHL scheme, with its $1/v_\infty^3$ dependence, does not have this issue. In fact, $v_\infty$ decreases with distance from the BH and simulation resolution, which should help promote accretion, though it is countered by the steeper drop-off in density (see top panels of Fig. 5 and Fig. 8), which suppresses accretion in the lower resolution simulations. This emphasises the issue of applying the mass-flux scheme in simulations that are too underresolved. While it can be difficult to ascertain what 'too underresolved' means for a given set-up, the inconsistency in the growth pattern here is delineated by marginally resolving the initial $R_{HL}$. This suggests that $R_{HL}$ could be a good indicator of the refinement required to appropriately use mass-flux to approximate BH accretion. Due to the higher initial $R_{HL}$ measurement in Halo 2, marginal $R_{HL}$ resolution was the maximum possible within reasonable computational expense. In line with Halo 1, the mass-flux scheme leads to progressively more efficient accretion as resolution decreases. The BHL scheme, however, does not exhibit a consistent trend; `2S.bf16` grows more rapidly than `2S.bf8` and `2S.bf4`, though still less than `2S.b01`. While the lack of growth with the BHL model at low resolution in Halo 1 gives weight to the argument that an underresolved ISM leads to an underestimated accretion rate, the fact that `2S.bf16` has a higher accretion rate than two simulations at lower resolution cautions against the use of this assumption in all cases.

## 4 DISCUSSION

### 4.1 Comparison with B18b

As mentioned in the introduction, this work builds upon the resolution study in B18b. We now compare our results with theirs, focusing on two sets of very similar simulations: `D_l26_tiny` and `R_128` and `1B.b01` and `1S.b01`. See Table 6 for a summary of key attributes. Their commonalities include BH mass, accretion scheme, and spatial resolution, while their differences lie in the host halo mass, available cooling channels (as discussed in Section 3.1.2), the minimum size of the high-resolution region, and the use of a drag-force algorithm. It is worth emphasizing that the precise implementation of our BHL models are extremely similar; gas properties are mass- and Gaussian-kernel-weighted, the kernel radii both have a max of $r_K = 2dx$, mass removal is limited to 75 per cent of the cell mass and there is no

**Table 5.** Summary of simulations of the $10.8\,M_\odot$ BH across two sets of initial conditions ('1S' and '2S') and accretion schemes ('b' for 'BHL' or 'm' for 'Mass-Flux') at various levels of refinement. From left to right, the columns represent (1) the simulation name; (2) the mass of the BH in solar masses; (3) the accretion scheme used, which can be either the BHL scheme (equation 1) or mass-flux (equation 11); (4) the number of cells by which the initial scale radius is resolved; (5) the resolution level of the ENZO simulation; (6) the minimum cell width in units of proper parsecs; and (7) the size of the high-resolution accretion region in number of cells.

| | | Post-Type II supernova BH simulations | | | | |
|---|---|---|---|---|---|---|
| Name | Mass [$M_\odot$] | Accretion scheme | Refinement factor | Level | d$x$ [pc] | $N_\mathrm{grid}$ [cells] |
| 1S.bf8 | 10.8 | BHL | 1/8 | 13 | $2.45 \times 10^{-02}$ | $5^3$ |
| 1S.bf4 | 10.8 | BHL | 1/4 | 14 | $1.23 \times 10^{-02}$ | $5^3$ |
| 1S.b01 | 10.8 | BHL | 1 | 16 | $3.07 \times 10^{-03}$ | $5^3$ |
| 1S.b04 | 10.8 | BHL | 4 | 18 | $7.69 \times 10^{-04}$ | $6^3$ |
| 1S.b04-no-SN | 10.8 | BHL | 4 | 18 | $7.69 \times 10^{-04}$ | $6^3$ |
| 1S.b08 | 10.8 | BHL | 8 | 19 | $3.84 \times 10^{-04}$ | $11$ |
| 1S.mf8 | 10.8 | Mass-Flux | 1/8 | 13 | $2.45 \times 10^{-02}$ | $5^3$ |
| 1S.mf4 | 10.8 | Mass-Flux | 1/4 | 14 | $1.23 \times 10^{-02}$ | $5^3$ |
| 1S.m01 | 10.8 | Mass-Flux | 1 | 16 | $3.07 \times 10^{-03}$ | $5^3$ |
| 1S.m04 | 10.8 | Mass-Flux | 4 | 18 | $7.69 \times 10^{-04}$ | $6^3$ |
| 1S.m04-no-SN | 10.8 | Mass-Flux | 4 | 18 | $7.69 \times 10^{-04}$ | $6^3$ |
| 1S.m08 | 10.8 | Mass-Flux | 8 | 19 | $3.84 \times 10^{-04}$ | $11^3$ |
| 2S.bf16 | 10.8 | BHL | 1/16 | 17 | $2.08 \times 10^{-03}$ | $5^3$ |
| 2S.bf08 | 10.8 | BHL | 1/8 | 18 | $1.30 \times 10^{-03}$ | $5^3$ |
| 2S.bf04 | 10.8 | BHL | 1/4 | 19 | $5.19 \times 10^{-04}$ | $5^3$ |
| 2S.bf4-no-SN | 10.8 | BHL | 1/4 | 19 | $5.19 \times 10^{-04}$ | $5^3$ |
| 2S.b01 | 10.8 | BHL | 1 | 21 | $1.30 \times 10^{-04}$ | $5^3$ |
| 2S.b01-no-SN | 10.8 | BHL | 1/4 | 19 | $5.19 \times 10^{-04}$ | $5^3$ |
| 2S.mf16 | 10.8 | Mass-Flux | 1/16 | 17 | $2.08 \times 10^{-03}$ | $5^3$ |
| 2S.mf8 | 10.8 | Mass-Flux | 1/8 | 18 | $1.30 \times 10^{-03}$ | $5^3$ |
| 2S.mf4 | 10.8 | Mass-Flux | 1/4 | 19 | $5.19 \times 10^{-04}$ | $5^3$ |
| 2S.mf4-no-SN | 10.8 | Mass-Flux | 1/4 | 19 | $5.19 \times 10^{-04}$ | $5^3$ |
| 2S.m01 | 10.8 | Mass-Flux | 1 | 21 | $1.30 \times 10^{-04}$ | $5^3$ |
| 2S.m01-no-SN | 10.8 | Mass-Flux | 1 | 21 | $1.30 \times 10^{-04}$ | $5^3$ |

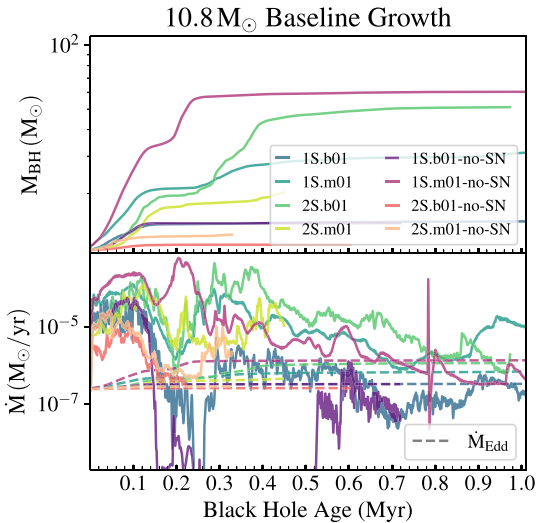

**Figure 11.** BH growth (top panel) and accretion rate (lower panel) of the $10.8\,M_\odot$ BH over 1 Myr for the baseline resolution group of simulations in each halo. Both the simulations with and without supernova feedback (-no-SN suffix) are included. The lower panel also includes the Eddington accretion rates as *dashed lines*. In contrast to the $270\,M_\odot$ BH, this smaller BH does not accrete the $600\,M_\odot$ cloud core.

Eddington cap on accretion. They are not identical because B18b use cloud particles to sample the local environment, whereas we take an average over all cells within $r_K$. This minor difference is unlikely to affect results.

The morphology of the gas in the B18b simulations broadly agrees with our findings. They too observe gas clumping as resolution increases and track the formation of a pc-scale slim 'nuclear' disc, despite the differences in cooling and initial conditions (though including feedback may disrupt this structure). We have already discussed in Section 3.2.2 that radiative cooling begins to dominate over advective cooling in the vicinity of the BH, as predicted in B18b. Additionally, most of their BHs also experience super-Eddington accretion that is smooth and continuous at low resolution and episodic at high resolution, reinforcing their conclusion that the degree of gas refinement has a significant impact on the accretion cycle.

However, there are some notable discrepancies in the behaviour of the BH particle. At almost precisely the same spatial resolution, our $270\,M_\odot$ BH (1B.b01) grows to $\simeq 2000\,M_\odot$ in 1 Myr, whereas the B18b $260\,M_\odot$ BH (D_l26_tiny) fails to grow at all. Density projections reveal that it oscillates unphysically in the direction perpendicular to the plane of a self-gravitating disc of gas it had become detached from during collapse (fig. 7 of B18b). The accretion rate varies by up to a factor of $10^6$ in less than 10 kyr, but is still kept low by extremely high relative velocities $v_\infty > 100\,\mathrm{km\,s^{-1}}$, about 100 times the values measured in 1B.b01. This, in turn, causes the $R_\mathrm{HL}$ in D_l26_tiny to be consistently underresolved by a large magnitude, whereas our simulation fluctuates between a few grid cells of refinement and underrefinement (see the final panel of Fig. 5a). The authors attribute this strange behaviour to $R_\mathrm{HL}$ not being resolved when the BH begins to accrete. They argue that $R_\mathrm{HL}$ resolution is a good proxy for drag-force resolution; when resolved i.e. $R_\mathrm{HL} > \mathrm{d}x$, the BH transfers momentum to the gas and decreases its velocity, causing $R_\mathrm{HL}$ to reduce further. When $R_\mathrm{HL} < \mathrm{d}x$ initially, the lack of dynamical friction causes the particle to become unbound from the gas and exhibit unphysical behaviour. Indeed, we set the refinement level in `1B.b01` such that $dx \simeq R_{HL}$ when the BH starts accreting, which just happened to be very close to the level chosen arbitrarily in B18b. The initial value of $R_{HL}$ in B18b is likely to be much smaller than our $R_{HL} = 0.012$ pc based on the $v_\bullet$ rate in the fourth panel of fig. 6 in B18b, and the fact that their host halo is $4 \times 10^5$ times the mass of our Halo 1. Their initial conditions drive up the resolution needed to resolve the dynamical friction between the gas and central stellar-mass BHs.

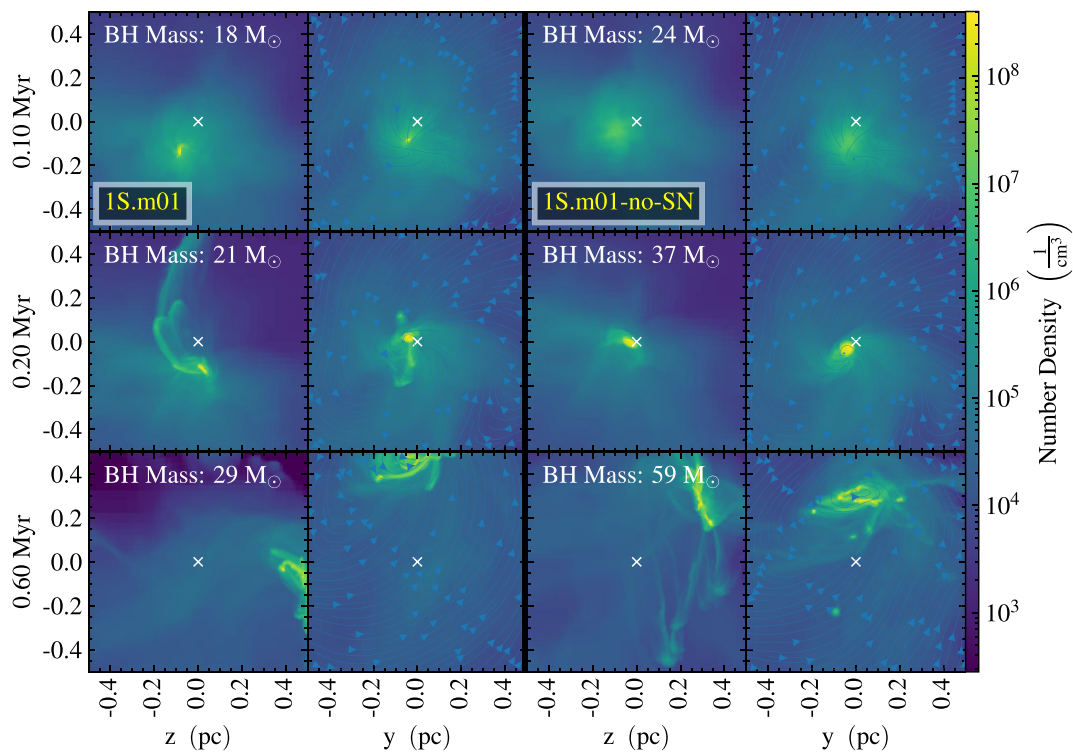

**Figure 12.** Total hydrogen nuclei number density projections face-on (first/third columns) and edge-on (second/fourth columns) of the collapsing cloud in `1S.m01` and `1S.m01-no-SN`. Time increases from top to bottom in variable increments, as shown on the *y*-axis. The BH location is marked by a *white cross* in each panel and the growing BH mass is indicated in the first column. Gas velocity streamlines are overlaid in blue on the edge-on projections to show the flow of gas towards a centre of gravity offset from the BH, which becomes a self-gravitating disc by $t = 0.6$ Myr. The gas in the simulation with no supernova event collapses more slowly and closer to the particle than its counterpart with the supernova, allowing the BH to accrete more mass before the disc drifts away.

To address the issue of the oscillating, accretion-less BH, B18b introduce a drag-force sub-grid model that involves forcing the particle to remain at the centre of the gas cloud until $R_{HL} > 0.2dx$. This is applied to a variety of masses, including a $10\,M_\odot$ BH in R_128. In all instances, highly efficient accretion to order $\sim 10^6\,M_\odot$ is observed. From fig. 13 in B18b, we can see that this is approximately equal to the mass enclosed within $r \simeq 1$ pc around the BH (albeit after 1.6 Myr of evolution), where $r \simeq 1$ pc corresponds to the beginning of the flattened core in the gas density profile, as seen in fig. 14. Likewise, our `1B.b01` BH quickly grows to $\sim 10^3\,M_\odot$, the order of magnitude of mass enclosed within the $\sim 0.1$ pc core as deduced from Fig. 1.

Conversely, our `1S.b01-no-SN` simulation, with almost identical BH mass and spatial resolution to R_128, grows by a meagre $5\,M_\odot$ in the first 0.15 Myr before plateauing. It does not display the same kind of oscillatory motion nor high relative velocities as `D_126_tiny`, though a self-gravitating disc does form very close by and drifts away over time. As discussed in Section 3.4, the mass-flux accretion scheme allows `1S.m01-no-SN` to accrete more from the gas cloud as it collapses, despite the particle not being located at the centre of mass, though not enough for it to capture the gas cloud into its orbit. Even as resolution is increased to resolving the initial $R_{HL}$ by 4 cells, `1S.b04-no-SN` shows little growth. The only high-resolution $10.8\,M_\odot$ BH run that accretes the $10^3\,M_\odot$ core is `1S.m04-no-SN`, as discussed in Section 3.4.2. Like `1S.b01`, the $R_{HL}$ resolution in `1S.b01-no-SN` varies mildly between $0.1dx$ and $10dx$ (see bottom panel of Fig. 13a), hence this simulation would rarely qualify for the application of the drag force algorithm used in B18b. Moreover, the authors caution that B18a shows that applying such a model when the drag force is well resolved could result in artificial acceleration of the BH particle. Thus, if we assume that the dynamical friction of the gas is, on average, resolved for `1S.b01-no-SN`, then unresolved gas dynamics cannot be responsible for its lack of growth. The fact that one simulation out of the 12 $10.8\,M_\odot$ BHs with 'marginal' to 'modest' $R_{HL}$-resolution runs grew to the order of the larger seed indicates that randomness of the small-scale dynamics can significantly influence the short-term growth of less massive seeds. Since most of these BHs are unlikely to accrete much past the 1 Myr cut-off anyway, as the self-gravitating gas has settled into a disc and moved away, it would take the introduction of new gas from a merger to trigger further growth in this BH.

In contrast to the $10.8\,M_\odot$ BHs, the $270\,M_\odot$ BHs in both haloes consistently have enough gravity at marginal $R_{HL}$ resolution to pull the gas into orbit and feed from it. These findings suggest that, when the relevant gas dynamics are resolved, larger seeds grow more reliably than smaller ones. This work has shown that a BH of mass $M_{BH} \sim 10^2\,M_\odot$ can reliably accrete in a dark matter mini-halo in the absence of feedback, whereas one of M$_{BH}$ ∼ 10$^1$ M$_\odot$ order grows only some of the time.

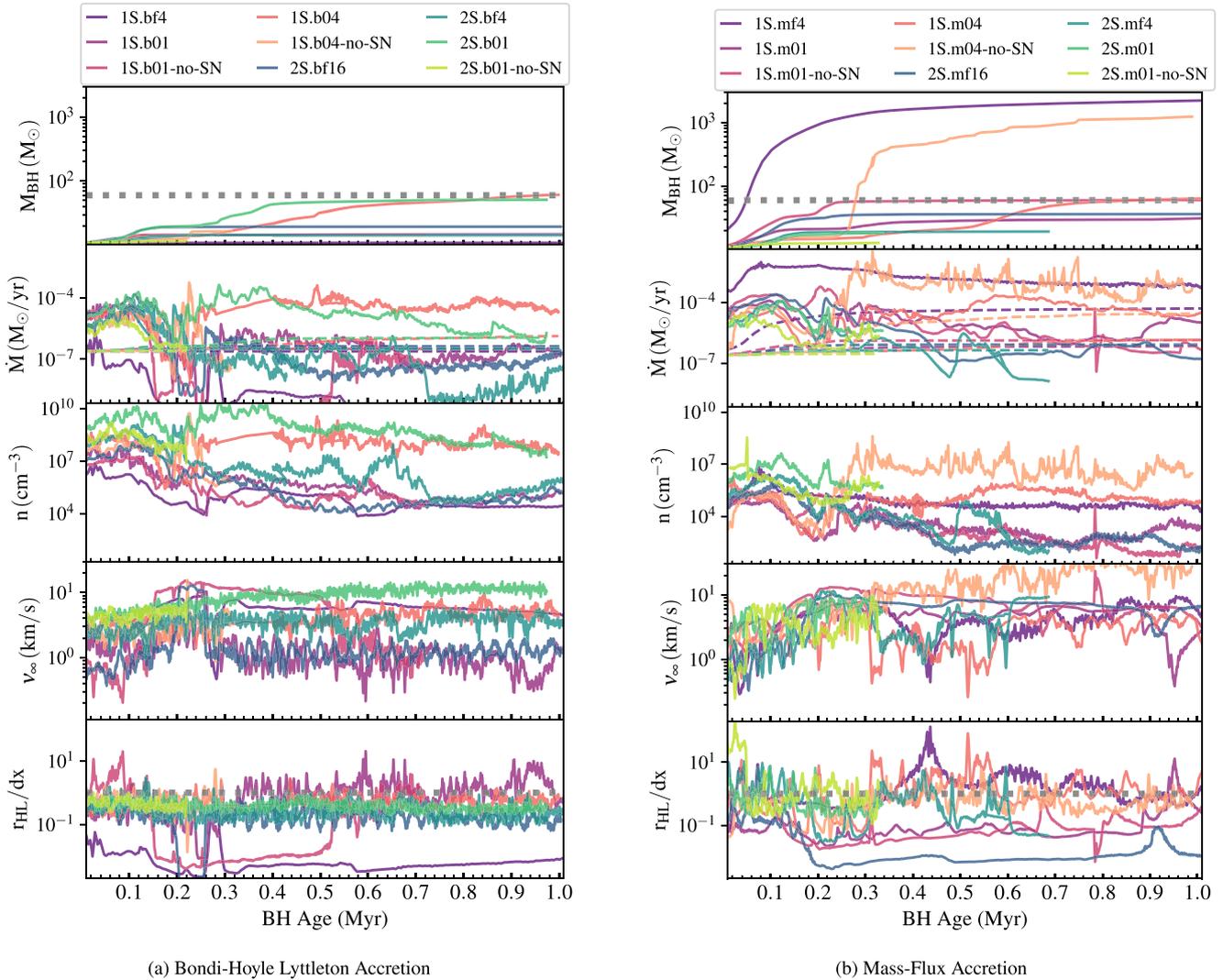

**Figure 13.** Time evolution of mass-averaged quantities in the accretion region of the 10.8 M$_\odot$ BH growing via the BHL scheme (left) and the mass-flux scheme (right). The four simulations in each sub-plot are identical except for their spatial resolution; `1S.xf8` has the lowest, `1S.x01` the highest (see Table 5 for further details). From top to bottom, the panels illustrate the mass of the BH, the accretion rate of the BH, the hydrogen nuclei number density of the gas, the velocity $v$ of the gas with respect to the BH, and the instantaneous Hoyle–Lyttleton radius resolution $r_{HL}/dx$, where d$x$ is the cell width of accretion region. The second panel includes the Eddington rate $\dot{M}_{Edd}$ as a *dashed line*. The data have been time-averaged and interpolated so that an equal number of points are plotted for each line.

The BH refinement scheme used in B18b differs from the scheme used in this work in a few key aspects, which may have impacted results. First, B18b gradually increases the refinement level around the BH over ≃50 kyr, as seen in the projected density maps in fig. 7 of B18b, whereas we reached the target resolution in just one timestep. Since the resolution at the time of BH formation is already quite high in our haloes (1.2 × 10$^{-2}$–8.3 × 10$^{-3}$ pc), the maximum number of levels jumped in one time-step is six (`2S.b01`) and the minimum zero. The `1S.b01-no-SN` simulation discussed in the previous paragraph only required one additional level of refinement when it began accreting, so the refinement scheme is unlikely to have impacted these and other results from simulations at a similar level. Additionally, no strange numerical effects (sharp boundary edges, anomalous structures) were apparent in any of the density projections of runs with significant refinement level jumps. Another difference between our scheme and B18b's lies in the size of the high-resolution region. The particle host grid in B18b is maintained at a volume of 8$^3$ cell widths ($n_{zoom}$ = 8). We impose a minimum volume of 5$^3$ cell widths on our high-resolution region, but in practice, it is usually much larger. This is evidenced by Fig. 2 in which the highest resolution grids extend to $r \simeq 0.1$ pc, just under 10 times the mandated $R_{HL}$ = 1.2 × 10$^{-2}$ pc. Any one of the four conditions detailed in Section 2.3 can trigger refinement, so even if the HL radius is well resolved, the local Jeans length, dark matter, or gas may not, and this results in a larger high-resolution region than anticipated by $R_{HL}$ refinement alone. This leads to the formation of dense gas outside of the BH accretion zone, which can fragment and self-gravitate. Some of our simulations reveal that the self-interaction of dense gas can trigger the expulsion of clumps from the accretion disc before they are absorbed by the BH. While in general, this may lead to enhanced BH growth on longer scales for simulations with small high-resolution grids, B18b include a star formation prescription that

**Table 6.** Summary of the key simulation parameters in `1B.b01` and `1S.b01` from this work and `D_l26_tiny` and `R_l28` from table 2 in Beckmann, Devriendt & Slyz (2018b). Note that the choice of d$x$ in `R_l28` is based on $R_{Bondi} \simeq dx$, which is likely to be smaller than $R_{HL}$.

|  | Comparison with B18b | | | |
|---|---|---|---|---|
|  | 1B.b01 | D_l26_tiny | 1S.b01-no-SN | R_l28 |
| ICs | Cosmo. | Galaxy | Cosmo. | Galaxy |
| Cooling | Grackle | Analytic | Grackle | Analytic |
| $M_{BH}$ [M$_\odot$] | 270 | 260 | 10.8 | 10 |
| $t_{form}$ [Myr] | 124.76 | 100.4 | 124.76 | 100.4 |
| d$x$ [pc] | $1.2 \times 10^{-2}$ | $1 \times 10^{-2}$ | $3 \times 10^{-3}$ | $4 \times 10^{-3}$ |
| $2r_{200}$ Mass [M$_\odot$] | $1.3 \times 10^7$ | $5.1 \times 10^{11}$ | $1.3 \times 10^7$ | $5.1 \times 10^{11}$ |
| $R_{HL}$ [pc] | $1.38 \times 10^{-02}$ | – | $3 \times 10^{-03}$ | $< 4 \times 10^{-3}$ |
| Accretion Scheme | BHL | BHL | BHL | BHL |
| Drag Force | – | – | – | Max[a] |
| $n_{zoom}$ | > 5 | 8 | > 5 | 8 |

*Note.* [a]B18b forces the relative velocity between the BH and the gas to be zero until $R_{HL}$ is resolved to d$x$/5. This constitutes a 'maximum' drag-force algorithm.

also limits the number of clumps being accreted. We conclude that the discrepancies between our refinement schemes are not likely to have caused significant differences in the outcome of the simulations.

Our study has supported the following findings of B18b: the formation of a 'nuclear' pc-scale disc that feeds the BH, the initial rapid growth in mass corresponds approximately to the cloud core mass, gas clumpiness at high resolution that leads to more episodic accretion patterns, and convergence of mass-flux and BHL schemes at high resolution. We also find evidence for the B18b prediction that radiative cooling dominates over advective cooling as resolution increases and three-body H$_2$ formation kicks in. Our explicit refinement of the scales relevant to accretion allowed us to investigate the growth of these small seed BHs without the need for a drag-force algorithm. Unlike B18b, our 10.8 M$_\odot$ BH does not grow to the magnitude of the core mass in most cases as it became very sensitive to local dynamics in both sets of accretion schemes, even though the dynamical friction force remained reasonably resolved.

### 4.2 Comparison with other works

Similarly to B18a, Negri & Volonteri (2017) found that the Bondi model converged with the local mass-flux value (measured within $R_{Bondi}$) as the radius within which the input gas attributes $c_{s,\infty}$ and $v_\infty$ were measured decreased. This is broadly supported by our simulations regardless of BH seed mass, with the exception of `1S.b04-no-SN` and `1S.m04-no-SN`. While `2S.b01` and `2S.m01` do not converge, they are only marginally resolving the $R_{HL}$, and limited resources did not allow the probing of higher resolutions for this set of simulations.

However, the conclusion of Negri & Volonteri (2017) that the mass-weighted Bondi model at low resolution and without feedback underestimates the accretion rate with respect to the mass-flux measured within $R_{Bondi}$ in a high-resolution simulation is not strongly supported by our findings using the full BHL model. For the 270 M$_\odot$ BH in Halo 1, `1B.b01` and `1B.m16` are both at $\simeq 2000$ M$_\odot$ after 1 Myr, while in Halo 2, `2B.b01` actually becomes $\simeq 600$ M$_\odot$ more massive than `2B.m08`. The smaller 10.8 M$_\odot$ in Halo 1 shows some evidence for the Negri & Volonteri (2017) finding. The simulation `1S.m04-no-SN` grows far more efficiently than `1S.b01-no-SN` and `1S.bf4`, with a final mass two orders of magnitude greater. However, in Halo 2, `2S.bf16` and `2S.m01` both grow by just a few solar masses. Since we have used $R_{HL}$ to measure the degree of resolution and Negri & Volonteri (2017) used $R_{Bondi}$, our 'marginally resolved' `b01` runs may have been considered more highly resolved in their study. Indeed, even our 'low-resolution' simulations resolved the Bondi radius well most of the time. We also observed that the mass-flux can sometimes overestimate the accretion rate if not resolved sufficiently (e.g., `1S.mf4`). None the less, the fact that we observed inconsistent trends in growth as resolution lowered cautions against the use of a boost factor, particularly a non-parametrised term.

All simulations in this work were at the same resolution level before the BH particle was inserted. The instruction to resolve the Hoyle–Lyttleton radius triggers instantaneous refinement by up to six levels in the central region of the collapsing gas cloud and, as we have discussed, there can be significant variations in the evolution therein. However, increasing the resolution of the halo before BH formation could have exacerbated these differences. While we have not performed resolution tests of this kind, Regan & Haehnelt (2009) found no systematic difference in the initial angular momentum of their atomic-cooling haloes when they increased and decreased the initial refinement level relative to their fiducial runs. They note that the most obvious dynamic variations emerge in the later stages of collapse, which corresponds to the branching stage of the simulations presented here. We have gauged this by comparing our halo density profile in Fig. 1 with those in fig. 5 of Regan & Haehnelt (2009).

### 4.3 Caveats

#### 4.3.1 Feedback

The BHs in this work originate from Pop III stars that lived between 2 and 3.8 Myr prior to gravitational collapse (Woosley, Heger & Weaver 2002), though we do not simulate their main-sequence lifetime. Ionizing radiation has been shown to efficiently evacuate gas from the centre of the mini-halo, either dispersing it to the outskirts or destroying the gas structure completely (Shapiro, Iliev & Raga 2004; Alvarez, Bromm & Shapiro 2006; Whalen et al. 2008; Latif, Whalen & Khochfar 2022). This would have significantly reduced the density of the gas in the vicinity of the particles, potentially limiting their initial growth to a small fraction of the Eddington rate (Alvarez, Wise & Abel 2009). However, Jaura et al. (2022) find that the ionizing radiation from Pop III stars does not escape the dense accretion disc surrounding them when photons are injected into the simulation on scales smaller than the local scale height of the disc. Consequently, the inclusion of radiative feedback has little impact on the total mass of protostars formed during the 20 kyr simulated. Therefore, the nature of the environment in which the first generation of stellar-collapse BHs are born – whether gas-rich or gas-poor – remains uncertain.

Moreover, the absence of BH feedback leads to overestimation of the true accretion rate. We did not impose an Eddington limit cap and as a result, most BHs grew at $10^3 \dot{M}_{Edd}$ on average initially. In the context of a resolution study, AGN feedback has been shown to increase in efficiency close to the BH as the ISM becomes more resolved, stifling the local mass-flux rate in comparison to a BHL scheme with input parameters further from the particle (Negri & Volonteri 2017). Given that the simple feedback from a supernova blast through the dense environment in these simulations led to a large divergence in the accretion cycles of two high-resolution BHs that were identical in every other respect (`1S.m04` and `1S.m04-no-SN`), it is likely that continuous accretion feedback will produce even more varied results across resolution for the small seeds (e.g. Johnson et al. 2011). We plan to include such feedback in future work.

*4.3.2 Star formation*

This work only allowed a single Pop III star to form per mini-halo simulated. We suppressed the formation of Pop III binaries at inception and the formation of stellar clusters from dense fragmented clumps that developed after the BH started accreting. Earlier studies suggest that only one star formed per mini-halo (Abel, Bryan & Norman 2002), but the degree of fragmentation seen in more recent work indicates binary systems (Turk, Abel & O'Shea 2009; Jeon et al. 2015) or stellar clusters may be more common (Regan, Johansson & Haehnelt 2014; Latif, Whalen & Khochfar 2022). Should multiple stars have been allowed to form and stellar feedback included, the gas reservoir feeding the BHs in work may have been reduced enough to severely limit the accretion, as in Smith et al. (2018), in which a population of stellar mass BHs from the RENAISSANCE suite is analysed. However, their resolution is much lower than ours, reaching a maximum of $dx = 1.2$ pc at $z = 15$ and $M_{DM} \sim 10^4 M_\odot$, and therefore do not resolve the Bondi radius. Along with underresolving dynamical friction, the kind of fragmentation which led to the spurious efficient growth of one of our $10.8 M_\odot$ BHs would likely not have been possible.

*4.3.3 Mass-flux and resolution*

It is worth emphasizing that although these simulations are considered 'high-resolution' by the standards of cosmological and galaxy simulations, they are not high-resolution in the absolute sense of resolving the physical accretion region of stellar-mass BHs. The Schwarzschild radii of 10.8 and $270 M_\odot$ BHs are 31.9 and 794.4 km, respectively, $\sim 10^{-8}$ times the minimum cell width reached in this work. Therefore, the pc-scale disc that developed in many of the $270 M_\odot$ simulations is far from the km-scale disc directly feeding the BH. The mass-flux across the inner radius of this disc would represent the most realistic estimate of the accretion rate, but this lies far beyond the physics included in cosmological codes and within the realm of GRMHD (General Relativistic Magnetohydrodynamics) simulations (Chatterjee et al. 2020). As such, measuring the mass-flux across a boundary so far from the event horizon, as we do in this work, constitutes a very crude approximation of BH accretion, subject to large-scale turbulence and other effects that would not influence the true accretion disc. As we have shown, for instance in Fig. 13(b), using mass-flux as an accretion scheme at too low a resolution and too far a distance from the BH will produce unrealistic results.

## 5 SUMMARY AND CONCLUSIONS

We have presented a resolution study of accretion on to 10.8 and $270 M_\odot$ black holes formed from Pop III stars in high-redshift mini-haloes. The simulations were evolved for 1 Myr, the approximate free-fall time of the cloud core in both haloes. Our high-resolution simulations endeavoured to resolve the Hoyle–Lyttleton radius $R_{HL}$ as measured when the black hole started accreting. In supersonic gas, this is purported to be the relevant scale for accretion physics (Bondi 1952) and potentially a good indicator of dynamical friction resolution (Beckmann, Devriendt & Slyz 2018b). The explicit focus on stellar-mass seeds, the inclusion of the three-body $H_2$ formation channel, and the use of two sets of cosmological initial conditions distinguish this study from previous works which tended to use exclusively massive seeds, less extensive primordial chemistry networks, and more idealized initial gas configurations.

Our main findings are summarised as follows:

(1) Accretion rates produced by the BHL model and a simple mass flux scheme converge to similar values in the majority of simulations as resolution increases.

(2) The accretion cycle becomes more episodic and less smooth as resolution increases due to a greater degree of gas fragmentation at high densities. This phase change is induced by the cooling length being more highly resolved.

(3) A rotationally supported disc develops around the $270 M_\odot$ black hole irrespective of resolution and accretion scheme, though it forms more quickly at high resolution. The disc acts as a mass reservoir for the black hole, enabling it to efficiently accrete most of the dense core of the collapsing gas cloud.

(4) Radiative cooling begins to dominate over advective cooling in the accretion region as resolution increases. The increased production of $H_2$ at higher gas densities facilitates this transition.

(5) More complex instabilities are seen in discs below a spatial resolution threshold of $dx \simeq 1 \times 10^{-3}$ pc. Bar-like overdensities form in the accretion region, sourcing spiral arms and redistributing angular momentum throughout the disc. Generally, the appearance of the bar correlates with rising accretion rates.

(6) The $10.8 M_\odot$ BH cannot attract enough mass to form a disc in most high- and low-resolution simulations. A self-gravitating disc grows independently nearby as the gas cloud continues to collapse offset from the BH position. Clumps are ejected from this disc, some of which are directed towards the particle and accreted. This unreliable source of gas results in modest growth, with a final mass of $M_{BH} < 60 M_\odot$ in all bar one simulation. While we did not include a star formation prescription, the density reached by many of these clumps makes them viable star candidates. Therefore, there is greater potential for the formation of a binary companion or cluster of stars around this $10.8 M_\odot$ black hole in comparison to the $270 M_\odot$ black hole.

In a broader context, our findings suggest that the use of a constant boost factor at low resolution would not lead to closer agreement in growth with high-resolution simulations. As more of the ISM is resolved, complicated gas dynamics emerge that introduce a degree of randomness in the evolution of the black hole particle. It is difficult to encompass the high-resolution behaviour in a subgrid model for use in a lower resolution simulation due to the non-linear nature of this phase transition. This is exemplified by the anomalous efficient growth of `1S.m04-no-SN` triggered by a collision with a wandering clump. Such random occurrences and their influence on accretion are challenging to incorporate in a model without simulating a wider array of seed stellar-mass BHs in mini-haloes to ascertain a 'typical' accretion rate. While this investigation is outside the remit of this study, it offers an interesting avenue for future research.

Moreover, the inclusion of feedback, both from the radiating Pop III progenitor and in the accretion process, would likely have a significant impact on the gas morphology and growth of these small BHs. In Alvarez, Wise & Abel (2009), a $100 M_\odot$ direct-collapse black hole formed in the wake of a radiating Population III star was found to barely grow both with and without accretion feedback. However, the resolution was $dx \simeq 0.03$ pc, of the order of some of the marginal to low-resolution simulations presented in this work. As we have seen, sub-$10^{-2}$ pc refinement can result in collisions with clumps and intense growth, completely altering the final mass reached on short time-scales. In future work, we will take advantage of our cosmological set-up to investigate accretion on longer time-scales and with feedback physics to establish the impact of resolution on stellar-mass BHs in a more realistic environment.

**ACKNOWLEDGEMENTS**

We would like to thank Ricarda Beckmann and Mike Petersen for their helpful discussions and insightful comments that enhanced the quality of this work. We also extend our gratitude to Molly Peeples for her generous provision of computing resources that enabled this work to be completed. SG is supported by the Science and Technologies Facilities Council (STFC) PhD studentship. BDS and SK are supported by the STFC Consolidated Grant RA5496. JR acknowledges support from the Irish Research Council Laureate programme under grant number IRCLA/2022/1165. JR also acknowledges support from the Royal Society and Science Foundation Ireland under grant number URF\R1\191132. The simulations were run on two high-performance computing (HPC) facilities, Cirrus and Pleiades. Cirrus is a UK National Tier-2 HPC Service at EPCC (http://www.cirrus.ac.uk), funded by the University of Edinburgh and EPSRC (EP/P020267/1). Pleiades is a distributed-memory SGI/HPE ICE cluster, part of the NASA High-End Computing (HEC) Program through the NASA Advanced Supercomputing (NAS) Division at Ames Research Center. Computations and associated analysis described in this work were performed using the publicly available ENZO code (Bryan & Enzo Collaboration 2014) and the YT (Turk et al. 2011) analysis toolkit, which are the products of collaborative efforts of many independent scientists from institutions around the world. For the purpose of open access, the authors have applied a Creative Commons Attribution (CC BY) licence to any Author Accepted Manuscript version arising from this submission.

**DATA AVAILABILITY**

All simulations in this work were run from a GitHub fork of ENZO (Bryan & Enzo Collaboration 2014) containing changes which, at the time of writing, have not been merged into the main branch. Details of the branch used are available on request from the author. All plots were produced using the YT Python package (Turk et al. 2011) and the scripts are available in this GitHub repository.

## APPENDIX A: THE RELATIONSHIP BETWEEN ACCRETION RATE AND SCALE RADIUS RESOLUTION

In this work, we set the spatial resolution of the simulation based on the measurement of the HL radius (equation 3) just before the BH starts accreting, as opposed to continuously resolving it. In Fig. A1, we examine the relationship between the real-time scale radius resolution (Hoyle–Lyttleton $R_{HL}$ and Bondi $R_{Bondi}$) and accretion rate for the 1B set of simulations. During the initial 100 kyr growth spurt (Fig. A1a), the relationship between $R_{HL}$ resolution and accretion rate is generally very weak. The near-zero Pearson coefficients $P$ of the aggregate of all simulations in each panel implies there is virtually no correlation between the two variables. Moreover, the very low $R^2$ values indicate that the relationship is not well approximated by the linear fit and the predictive power of radius resolution to determine the accretion rate is limited. Over the whole simulation period of 1 Myr (Fig. A1b), these trends largely persist, although the mass-flux scheme shows a slightly stronger negative linear relationship with $R_{Bondi}$ resolution, with $P = -0.56$ and $R^2 = 0.31$. Overall, improving the initial resolution of the HL radius from 1 to 16 cells (1B.m01 versus 1B.m16) does not significantly affect the accretion rate distribution nor how well the scale radii are continuously resolved, though there is some tendency for the upper limit of $R_{Bondi}$/dx to increase with resolution. This may be due to the fact that while the maximum relative velocity of the gas increases with simulation resolution (see the fourth panels of Fig. 5), sound speed remains comparatively stable (shown indirectly through the increasing circular Mach speed $v_\theta/c_s$ as simulation resolution increases in the fourth panels of Fig. 8).

In Fig. A2, we perform the same comparison for the 1S set of simulations. During the initial growth phase (Fig. A2a), the correlations are strong and positive for the BHL scheme, and weakly to moderately negative for the mass-flux scheme. This implies that as scale radius resolution increases in the BHL scheme, the accretion rate also increases, but the opposite is true for the mass-flux scheme. This trend is evident in Fig. 13, where the 1S.b04 and 1S.m04 accretion rates align well, but the lower resolution values are much smaller for the BHL scheme but similar to larger for the mass-flux scheme. As discussed in Section 3.4.2, this is due to the contrasting gas velocity dependence between the accretion models. It is also the case that the *spatial* resolution of the simulation correlates with the radius resolution (and thus the accretion rate) in the BHL scheme; the underresolved simulations 1S.bf4/8 do not exceed $R_{HL}/dx \simeq 0.1$ and $\dot{M}/\dot{M}_{Edd} \simeq 10 \, M_\odot$/yr, whereas the two highest resolution simulations 1S.b01/4 reach $R_{HL}/dx \simeq 10$ and $\dot{M}/\dot{M}_{Edd} \simeq 10^3 \, M_\odot$/yr. While this clustering is not as pronounced in the mass-flux case, the simulations with the highest spatial resolution tend to reach the highest scale radius resolutions and have a lower maximum accretion rate (1S.m04 and 1S.m01) do not exceed $\dot{M}/\dot{M}_{Edd} > 300 \, M_\odot$/yr, whereas 1S.mf4 and 1S.mf8 almost reach $\dot{M}/\dot{M}_{Edd} \simeq 1000 \, M_\odot$/yr. Past the initial growth phase (Fig. A2b), the mass-flux scheme exhibits increasing variance and both scale radius correlations with accretion rate decline to a virtually negligible value. The simulations using BHL accretion maintain their trends, though the strength of the linear relationship weakens to a moderate value, with $R^2 = 0.73$ and $R^2 = 0.51$ for $R_{HL}$ and $R_{Bondi}$ resolutions. respectively. The difference between not resolving and even marginally resolving the initial $R_{HL}$ when applying either scheme to the 10.8 $M_\odot$ BH is significant for the initial mass growth.

The Halo 2 simulations exhibit similar trends to the Halo 1 simulations in the 270 $M_\odot$ case, but since a greater degree of poor $R_{HL}$ resolution could be explored for the 10.8 $M_\odot$ BH, there is one noteworthy finding. For the BHL scheme, the lowest resolution simulation 2S.bf64 has the lowest $R_{HL}$/dx values, but the same $R_{Bondi}$/dx values as the highest resolution simulation 2S.b01. Since 2S.bf64 grew much less efficiently than 2S.b01, this suggests that merely resolving $R_{Bondi}$ is not a good indicator of resolved gas dynamics and will not lead to converged accretion rates as simulation resolution increases.

Overall, this analysis has shown that there is not a strong correlation between the continuous $R_{HL}$/dx resolution once $R_{HL}$ has been initially at least marginally resolved (with the caveat that we did not explore underresolving $R_{HL}$ for the 270 $M_\odot$ BH). However, underresolving the $R_{HL}$ radius does lead to contrasting accretion trends between the accretion schemes. Likewise, initially underresolving $R_{HL}$ generally leads to lower $R_{HL}$/dx values and correspondingly low accretion rates for the BHL scheme, whereas $R_{Bondi}$/dx values have been seen to increase again in very low-resolution simulations, though with no effect on the accretion rate. This suggests that $R_{HL}$ resolution is a more reliable measure of the resolution of the dynamics relevant to accretion than $R_{Bondi}$, even for mass flux accretion (as the mass flux and BHL rates converge at high resolution).

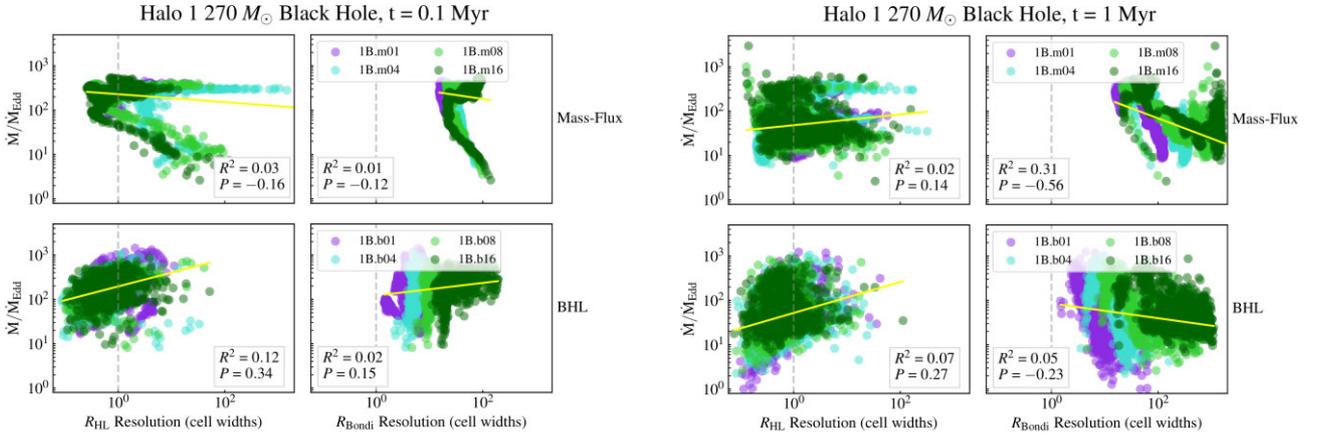

(a) Relationship between scale radius resolution and accretion rate during the initial growth phase (0.1 Myr) for the 1B set of simulations.

(b) Relationship between scale radius resolution and accretion rate over the entire period (1 Myr) for the 1B set of simulations.

**Figure A1.** Eddington-normalised accretion rate versus HL and Bondi radius resolution in units of cell widths for `1B` group of simulations for both accretion schemes during the first 0.1 Myr (left) and 1 Myr (right) of BH growth. The Pearson correlation coefficient $P$ and the coefficient of determination $R^2$ are calculated for the aggregate of data on the plot, independent of simulation. The *grey dash–dot line* indicates the boundary between an underresolved and resolved radius, i.e. $R_{\rm HL/Bondi}/{\rm d}x = 1$.

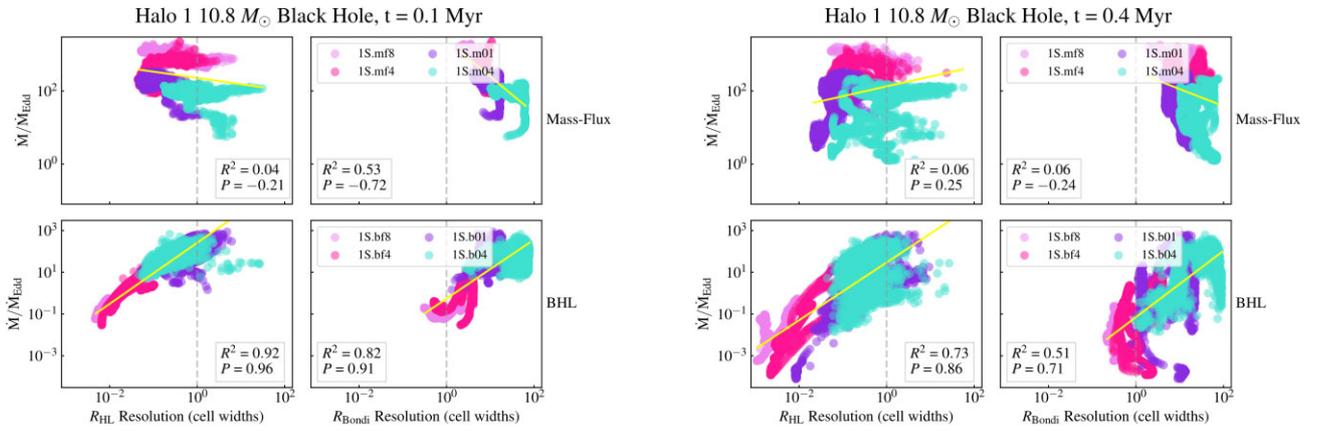

(a) Relationship between scale radius resolution and accretion rate during the initial growth phase (0.1 Myr) for the 1S set of simulations.

(b) Relationship between scale radius resolution and accretion rate for the entire duration (0.4 Myr) for the 1S set of simulations.

**Figure A2.** The same as Fig. A1 but for the `1S` set of simulations.

This paper has been typeset from a T<sub>E</sub>X/L<sup>A</sup>T<sub>E</sub>X file prepared by the author.